\documentclass{aastex}
\usepackage{emulateapj5}
\usepackage{apjfonts}
\usepackage{epsf}
\usepackage{lscape}

\def\cale{\ifmmode {\cal E} \else {cal E}\ \fi}
\def\ergs{$\rm erg~s^{-1}$}
\def\oiii{\ifmmode [O {\sc iii}] \else [O {\sc iii}]\ \fi}
\def\kms{\ifmmode {\rm km~s^{-1}} \else {\rm km~s^{-1}}\fi}
\def\mbh{M_{\rm BH}}

\def\sunm{M_{\odot}}

\begin{document}

\title{The Unified Model of Active Galactic Nuclei:  I. Non-hidden Broad Line Region Seyfert 2 and 
Narrow Line Seyfert 1 Galaxies}

\author{ En-Peng Zhang\altaffilmark{1,2} and Jian-Min Wang\altaffilmark{1,3}}

\altaffiltext{1}{Key Laboratory for Particle Astrophysics, Institute of
High Energy Physics, Chinese Academy of Sciences, Beijing 100039, China.} 
\altaffiltext{2}{Graduate School, Chinese Academy of Sciences, Beijing 100039, China}
\altaffiltext{3}{Corresponding author, wangjm@mail.ihep.ac.cn}

\slugcomment{Received 2006 April 4; accepted 2006 June 5}
%\slugcomment{received 2006 Apri l; accepted 2006??}
\shorttitle{Unified Model of AGNs}
\shortauthors{ZHANG AND WANG}

\begin{abstract}
The unified model of Seyfert galaxies suggests that there are hidden broad-line regions (HBLRs) in  
Seyfert 2 galaxies (S2s). However, there is increasing evidence for the appearance of a subclass of  
S2s lacking of HBLR (non-HBLR S2s). An interesting issue arises as to relations of non-HBLR S2s with 
other types of Seyfert galaxies and whether or not they can be included in the unified model. We 
assemble two sub-samples consisting of 42 non-HBLR S2s and 44 narrow-line Seyfert 1s (NLS1s) with 
redshift $z\le 0.05$ from published literatures to explore this issue. 
We compare black hole masses in the galactic centers, accretion rates, infrared color ratio 
($f_{60 \mu \rm m}/f_{25 \mu \rm m}$) as a potential indicator of the dusty torus orientation, 
\oiii $\lambda 5007$, radio and far infrared luminosities. We find that non-HBLR S2s 
and NLS1s have: 1) similar distributions of the black hole masses ($10^6-3\times 10^7\sunm$) and the
Eddington ratios ($L_{\rm Bol}/L_{\rm Edd}\sim 1$); 
2) significantly different distributions of $f_{60 \mu \rm m}/f_{25 \mu \rm m}$ ratios; 3) similar 
distributions of bulge magnitudes and luminosities  
of [O {\sc iii}], radio, far infrared emission. The similarities and differences can be 
understood naturally  if they are intrinsically same but non-HBLR S2s are viewed at larger angles 
of observer's sight than NLS1s. We thus suggest that non-HBLR S2s only have "narrower" broad line 
regions and they are the counterparts of NLS1s viewed at high inclination angles.
The absence of the polarized emission line in non-HBLR S2s
is caused by the less massive black holes and high accretion rate similar to NLS1s.
The implications of the unification scheme of non-HBLR S2s and NLS1s are discussed.
\end{abstract}
\keywords{galaxies: active --- galaxies: Seyfert --- polarization} 

\section{Introduction}
Seyfert galaxies were traditionally divided into two classes according 
to the presence or absence of the broad permitted optical lines. With the 
discovery of the polarized broad lines in NGC 1068, a prototypical Seyfert 2 galaxy,
Antonucci \& Miller (1985) suggested that there should be a geometrically 
and optically thick "torus" surrounding a Seyfert 1 nucleus with broad lines (BLS1s). 
If the torus is face on, we can "see" the broad line region directly and 
the galaxies appear as BLS1s. Otherwise it appears as a Seyfert 2. This is the basic 
idea of the unified model (UM) of AGNs (Antonucci 1993). Many 
authors have reported pieces of evidence for the orientation-based UM 
(Miller \& Goodrich 1990; Tran et al. 1992; Mulchaey et al. 1994, Tran 1995; Yong 
et al. 1996; Heisler, Lumsden \& Baily 1997, hereafter HLB97; Moran et al. 
2000; Lumsden et al. 2001). It has been found that at least $35\%$ of Seyfert 2 galaxies have
broad emission lines in polarized lights (Tran 2001; Moran et al. 2000) and $\sim 96\%$ of
the objects have column densities ranging from $10^{22}$ to $10^{24}$cm$^{-2}$ (Risaliti et al.
1999; Bassani et al. 1999). However, spectropolarimetric surveys of complete samples of Seyfert 2 
galaxies suggest that hidden Seyfert 1 nuclei have not been detected in $>50\%$ of these objects 
from the CfA and 12 $\mu$m samples of Seyfert 2 galaxies (Tran 2001, 2003); and $10-30\%$ S2 are 
found unabsorbed in X-rays (Panessa \& Bassini 2002), even $50\%$
among {\em ROSAT}-selected Seyferts (Gallo et al. 2006). Non-HBLR S2s are shown to be systematically 
weaker than their HBLR counterparts,
and can not be explained by different orientations (Tran 2001; 2003, Lumsden \& Alexander 2001),
challenging the unification scheme. Tran (2003) suggested that there are "true" Seyfert 2 galaxies.
The paradigm of the unification scheme for {\em all} 
Seyfert galaxies remains a matter as debate among the literatures
(e.g. Miller \& Goodrich 1990; Kay 1994; Tran 2001, 2003). 

We still poorly understand why we can not detect polarized broad lines (PBLs) in 
some S2s. The absence of PBLs could be  attributed to edge-on line of sight and 
hidden of electron scattering region (Miller \& Goodrich 1990; HLB97; Taniguchi \& Anabuki 1999).  
Nicastro (2000) related the absence of PBLs to higher Eddington ratios. He suggested 
that the width of broad emission lines is Keplerian velocity of an accretion disk at a
critical distance from the central black hole, which is the transition radius between
radiation  and gas pressure-dominated region (see also Nicastro et al. 2003). 
It has also been suggested that some non-HBLR S2s are intrinsically weak and lack of 
broad line region (Tran 2001, 2003; Gu \& Huang 2002; Laor 2003). As argued by Tran (2003),
"it appears that much of the difference between S1s and S2s can be explained solely by orientation,
it would be difficult for the same model to apply among the HBLR and non-HBLR S2s without
invoking intrinsic physical differences",
what are the physical meanings for the absence of polarized broad lines in Seyfert 2 galaxies?
Do they really not have "broad" line region?

It has been suggested that some of Seyfert 2 galaxies without PBLs as 
a new subclass are probably  lack of broad line region (Tran 2001, 2003). Tran (2003) 
confirmed that polarized hidden broad-line region S2s share many similar large-scale 
characteristics with BLS1s, while non-HBLR S2s do not. Deluit (2004) analyzed {\em Beppo}-SAX data
of Seyfert 2s and found that non-HBLR S2s are different in hard X-rays (15$-$136keV) from those with 
hidden BLR Seyfert 2. There is growing evidence for that not
{\em all} Seyfert 2 galaxies might be intrinsically similar in nature. As we discuss in detail in \S4
(see Table 1. for possible types of Seyfert 2 galaxies), 
some non-HBLR S2s may result from fuel-depleting Seyfert 1 and 2 galaxies if the dusty torus
is supplying matter onto the black holes (Krolik \& Begelman 1988). These objects 
could be characterized by low or absent absorption in X-ray band, they thus might be the progenitors
of the optically-selected unabsorbed Seyfert 2 galaxies defined by Panessa \& Bassini (2002)\footnote{The 
roles of the gas-dust ratio has been discussed by Maiolino et al. (2001) and Gallo et al. (2006)}.  
This definitely makes it more complicate to study the physics of non-HBLR S2s.  However, this paper focuses 
on the absorbed non-HBLR S2s. If they were powered by less massive black holes and obscured by
torus at larger viewing angles (Tran 2003), what are their counterparts at low orientation? This motivates 
us to explore the relation between non-HBLR S2s and narrow line Seyfert 1 galaxies.

{\footnotesize
\begin{center}{\sc Table 1 The possible types of Seyfert 2 galaxies}
\vglue 0.3cm
\begin{tabular}{lcc}\hline\hline
            &  absorbed                           & unabsorbed \\ 
            &  ($N_{\rm H}\ge 10^{22}$cm$^{-2}$)  & ($N_{\rm H}< 10^{22}$cm$^{-2}$)\\ \hline
non-HBLR    &  $\surd$                                    & $\surd$ (unabsorbed Seyfert 2)   \\
HBLR        &  $\surd$(classical Seyfert 2)               & $\surd$ (Gallo et al. 2006) \\ \hline 

\end{tabular}
\parbox{3.1in}
{\baselineskip 9pt
\noindent 
{\sc Note:} 
The symbol $\surd$ indicates that this type is generally
observed. Polarized spectroscopic measurements are available only for
four unabsorbed Seyfert 2 galaxies, NGC 2992 (Rix et al. 1990), NGC 5995 (Lumsden \& Alexander 2001, Tran 2001),
NGC 7590 (Heisler et al. 1997) and NGC 4501 (Tran 2003, Cappi et al. 2006). A polarized broad H$\alpha$ line has 
been found in the first two objects, but not in NGC 4501 and NGC 7590. Gallo et al. (2006) find $\sim 50\%$ of
{\em ROSAT}-selected Seyfert galaxies are low absorption Seyfert 2s.}  
\end{center}
}

As a distinct subclass, NLS1s have very narrow 
Balmer lines [FWHM (H$\beta$) $\le 2000~\rm km~ s^{-1}$], strong Fe {\sc ii} lines 
(Osterbrock \& Pogge 1985), and violent variability in soft X-ray band (Boller et al. 1996). 
They likely contain less massive black holes at the Eddington limit rates
(Boller et al. 1996; Laor et al. 1997) and can be explained by slim disk 
(Wang et al. 1999; Wang \& Zhou 1999, Mineshige et al. 
2000; Wang \& Netzer 2003; Wang 2003; Ohsuga et al. 2003, Chen \& Wang 2004; Collin et al. 2002; Collin 
\& Kawaguchi 2004, Kawaguchi et al. 2004). The brighter soft X-ray fluxes 
favor the pole-on orientation hypothesis since an edge-on thick disk is dimmer than the lower 
inclination (Madau 1988; Boller et al. 1996; Leighly 1999a, b; see more detail calculations of Watarai 
et al. 2005). It is thus expected that the soft X-ray selected NLS1s tend to have a pole-on orientation to 
observers (Boller et al. 1996). This is further supported by the polarization observations 
showing that most of the soft-X-ray-selected {\em ROSAT} AGNs have polarization lower than $\le 1\%$
and no clear optical reddening (Grupe et al. 1998a). Optically-selected NLS1s from the Sloan Digital Sky 
Survey (SDSS)  are weak in soft X-ray bands and hence have lower Eddington ratios (William et al. 2004).
If the dusty tori generally exist in Seyfert galaxies and their orientations  
are random,  what are the counterparts of NLS1s viewed at larger angles? The presence of the non-HBLR S2s 
and NLS1s as new members of Seyfert galaxies have strong impact on the classical unified model. 
We suggest that non-HBLR S2s are the counterpart of NLS1 viewed at
larger angles. If so, the black hole masses, accretion rates (also Eddington ratios) as fundamentally
intrinsic parameters should have same distributions.

In this paper, we compared large-scale properties of non-HBLR S2s and NLS1s to quantitatively 
test the above issues. We find non-HBLR S2s share the potential isotropic 
characters with NLS1s while they are different greatly in the potential 
anisotropic properties. This  may suggest that they are basically the 
same objects but viewed from different angles, adding new ingredient to the classical unification scheme. 

\section{Sample and Data} 
The main goal of the present paper is to show whether the absorbed non-HBLR S2s are intrinsically same with 
NLS1s but only viewed at high inclination angles. The first task for us is to define an homogeneous 
and complete sample for the two kinds of objects. However we encounter difficulties since the complete surveys 
of non-HBLR S2s and NLS1s in Seyfert galaxies were not made simultaneously. The current samples are  
insufficient and the fractions of NLS1s and non-HBLR S2s in Seyfert galaxies 
are uncertain. For non-HBLR S2, Moran et al. (2000) only give an upper limit fraction ($>75\%$) of Seyfert 2 
galaxies are non-HBLR S2s whereas Tran (2003) reported about $50\%$. For NLS1s, their fraction is $\sim 11\%$ in 
an optically-selected heterogeneous sample (Marziani et al. 2003), $\sim 15\%$ in SDSS sample (Williams et al. 
2002), which is much lower than $\sim (31-46\%)$ in soft X-ray selected Seyfert galaxies (Grupe et al. 1999; 
Grupe 2004 and Salvato et al. 2004). The best way for us is to collect 
{\em all} the available objects from the published literatures so that we can
avoid the absence of some non-HBLR S2s and NLS1s known.

We find about 150 NLS1s with available data and 46 non-HBLR S2s from published literatures as we can.
NLS1s are mainly from Boller et al. (1996), Veron-Cetty et al. (2001), Grupe et al. (1999; 2004), Stepanian 
et al. (2003) and Williams 
et al. (2002, 2004). Fig. 1 shows the redshift distributions of the total objects from the literatures.
The NLS1s have a wider redshift distribution (but $z\le 0.5$) whereas the non-HBLR S2s are less than $z\le 0.1$. 
It is not an easy job to perform spectropolarimetric  observations for identification of a non-HBLR S2.
The size of the present sample is thus limited. The non-HBLR S2s are mainly from HLB97, Lumsden et al. (2001), 
Tran (2003) and Lumsden et al. (2004). However as we argued
in \S1 and \S4, we exclude those non-HBLR S2s with $N_{\rm H}< 10^{22}$ cm$^{-2}$ if there are available data 
of X-ray observations. However, we are not able to guarantee the purity of the non-HBLR S2s with 
$N_{\rm H}\ge 10^{22}$ cm$^{-2}$ in our sample since some of non-HBLR S2s (20 out of 42 objects)
have no X-ray observations. The present sample thus might be mixed with those unabsorbed Seyfert 2galaxies 
(see \S4.1), which could have more massive black holes. To avoid luminosity selection bias due to the 
redshift difference when comparing their properties, we confine our sample within $z\le 0.05$ from 
non-HBLR S2s. There leaves a sample composing of 42 non-HBLS2s and 44 NLS1s. Table 2 and 3 list the two 
sub-samples and give observable and deduced parameters. We do not intend to select the soft X-ray steeper
NLS1s, but most of the present NLS1s have very 
steep soft X-ray spectra ($\Gamma_{\rm SX}>2.0$) as shown from Table 2 (only nine objects with $\Gamma_{\rm SX}<2$). 
This selection favors those NLS1s with pole-on orientations. We marked those objects with stars, which are
from 12 $\mu$m sample
so that we can compare the distributions of the present sample with 12 $\mu$m sample.
It should be noted that there are only 8 NLS1s in the 12 $\mu$m (IR selected) sample, which lacks NLS1s,
compared with soft X-ray selected samples.
Fortunately it is found that the present sample can well present the complete non-HBLR S2s in 12 $\mu$m
sample from the subsequent sections.

If NLS1s and non-HBLR S2s are basically the same objects but viewed from different angles, 
the observable properties related to the direction would be greatly different while those 
unrelated to the direction would be similar. The central engines can be described by three 
key parameters: the black hole mass, the accretion rate and the orientation of the torus. 
These parameters, as indirect observables, can be properly deduced from observations.
For Seyfert galaxies, the potential parameters of the isotropic properties are \oiii $\lambda 5007$ 
line emission, far-infrared continuum, and core radio continuum (Mulchaey et al. 1994).
\oiii line emission could be a good indicator of the ionizing luminosities.
Near/mid infrared emission is regarded as being anisotropic whereas far infrared is isotropic.
As for the potential anisotropic properties, we use the flux ratio of $f_{60\mu}/f_{25\mu}$ 
(hereafter $f_{60}/f_{25}$) as originally suggested by HLB97. We only use the core radio emission 
of Seyfert galaxies taking from the published literatures. Radio flux densities in Table 2 and 3 are 
at 5GHz or extrapolated to 5 GHz  assuming $F_{\nu} \propto \nu ^{-0.7}$ if the measurements 
are not at 5GHz. Most of the data were obtained by the VLA observation with a spatial resolution 
$0.1\sim 1.0^{\prime\prime}$, which corresponds to a size of $\sim 0.1-1$kpc for a redshift 
of $z\le 0.05$ in the present sample. \oiii luminosities are taken from the published literatures and
have been corrected for extinction. The infrared data are from {\it IRAS}. 
According to the reprocessing model, the distance emitting $\lambda$ photon
to the center is $27L_{\rm 45}^{1/2}\lambda_{60}^2=75L_{\rm 45}^{1/2}\lambda_{100}^2$pc, 
where $L_{\rm 45}=L/10^{45}$\ergs, $\lambda_{60}=\lambda/60\mu$m and $\lambda_{100}=\lambda/100\mu$m. 
The IR size is much smaller than  the radio region, however the region contaminated by star formation
may be comparable to the region observed by radio. We use the Hubble constant $H_0=75~\rm km~s^{-1}~Mpc^{-1}$ 
and deceleration factor $q_0=0.5$ throughout the paper.

\figurenum{1}
\centerline{\includegraphics[angle=-90,width=8.0cm]{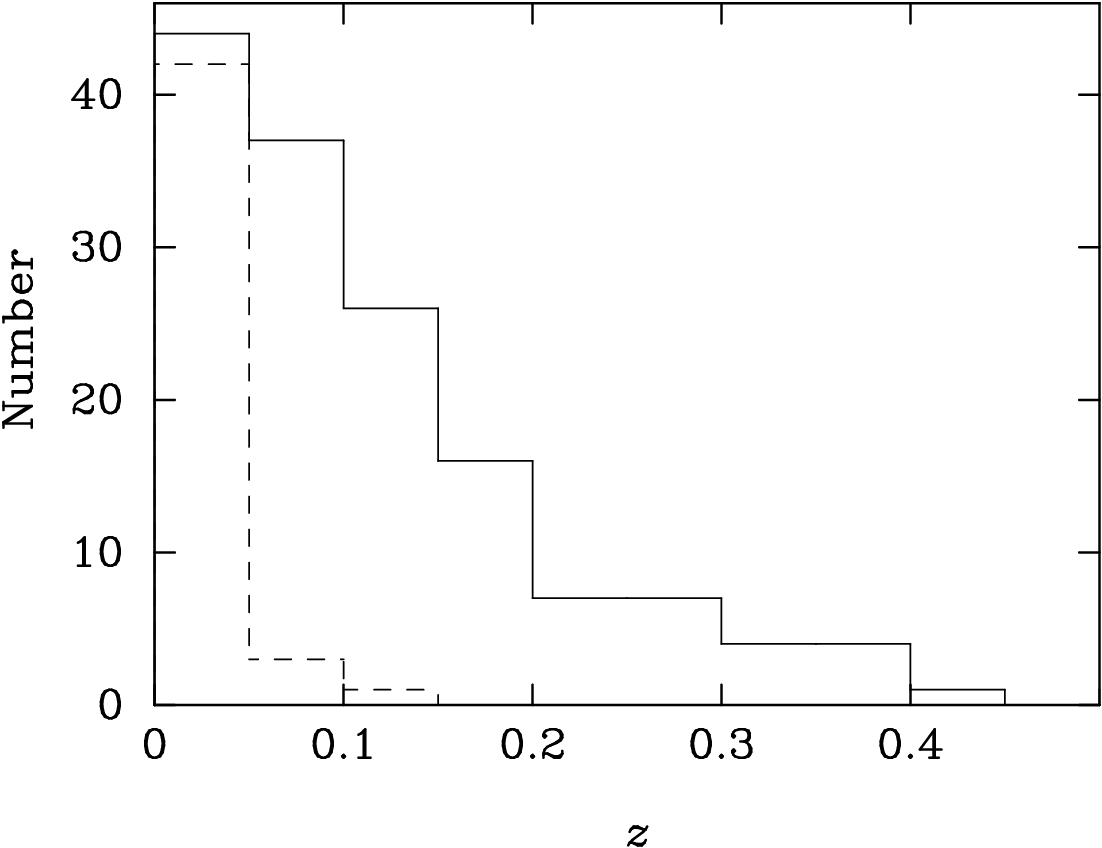}}
\figcaption{The redshift distribution of the present sample. 
solid line: NLS1s, the dashed line: non-HBLR S2s.}
\label{fig1}    
\vglue 0.3cm 

Finally we have to point out that the present sample is heterogeneous and incomplete, however,
 it represents the largest sample of non-HBLR S2 composed of all the known objects at $z\le 0.05$. 
We do not focus on the relative numbers of NLS1s and non-HBLR S2s, and the
relative number to their counterparts with broad lines, the conclusions in this paper might be
weakly dependent of  the heterogeneity of the present sample. Though the present sample covers
the 12 $\mu$m sample (Rush et al. 1993), as a comparison, we separately plot the later in order to 
carefully make conclusions in this paper. We stress that the data from 
the published literatures are from different authors and instruments. This leads to some uncertainties, 
but does not affect our conclusions within the error bars.

\section{Comparison of Properties}
\subsection{Black Hole Masses and Distributions} 
There are several independent methods to estimate the black hole masses in active galactic nuclei:
1) reverberation mapping (Netzer 1990, Netzer \& Peterson 1997, Kaspi et al. 2000); 2) empirical 
reverberation relation
(Vestergaard 2002, Kaspi et al. 2005); 3) $M_{\rm BH}-\sigma$ relation (Ferrarase \& Merrit 2000, 
Gebhardt et al. 2000a,b, Tremaine et al. 2002) or
$M_{\rm BH}-M_{\rm R}$ relation (McLure \& Dunlop 2001), where $\sigma$ is the velocity dispersion 
and $M_{\rm R}$ is the absolute magnitude of the host galaxies 
at $R-$band; 4) FWHM of \oiii as a substitute for the stellar velocity dispersion of galaxy bulges
(Nelson \& Whittle 1995; Boroson 2003; Shields et al. 2003; Greene \& Ho 2005). These methods allow
us to determine the black hole mass in different ways. For NLS1s, we can use both of the empirical
relation of the reverberation mapping and $M_{\rm BH}-\sigma/{\rm FWHM_{\rm [O III]}}$ relation.
In principle, we do not know whether the empirical reverberation relation is 
available or not in narrow line Seyfert 1 galaxies. Wang \& Lu (2001) applied the methods 2) and 4)
to NLS1s and estimated the masses of the black holes for a heterogeneous sample of Veron-Cetty
et al. (2001). They find the two methods are consistent within the uncertainty of 0.5 dex.
Shields et al. (2003) show this estimation is in agreement with that obtained from \oiii width. 
   
We estimate $R_{\rm BLR}$ through the empirical reverberation relation
(Kaspi et al. 2000), Vestergaard (2002) corrected BLR size-luminosity relationship as
\begin{equation}
R_{\rm BLR}=23.4\left(\frac{\lambda L_{5100}}{10^{44}{\rm erg~s^{-1}}}\right)^{0.56}{\rm  lt-days},
\end{equation}
for Seyfert galaxies and we  adopted it for our sample.  
The black hole mass can be obtained from $\mbh=v^2R_{\rm BLR}/G$, where $v=\sqrt{3}/2~{\rm FWHM}_{\rm H\beta}$ 
and $G$ is the gravitational constant, if the BLR size $R_{\rm BLR}$ is known from the
empirical relation or mapping technique\footnote{The zeropoint of the black hole mass from the reverberation 
mapping has been recalibrated by Onken et al. (2004) and Kaspi et al. (2005). The new calibration increases 
the zeropoint by roughly a factor of 2,
however, it has not been established with great statistical certainty (Nelson et al. 2004; Grenne \& Ho 2006).
Here we use the calibrated relation for Seyfert galaxies from Vestergaard (2002).}. 

However, for the non-HBLR S2s, the  empirical reverberation relation is not available since  the optical
continuum is strongly absorbed by the dusty torus. 
We estimate the black hole masses in  non-HBLR S2s  by the relation    
 $M_{\rm BH}=1.35\times10^{8}\left({\sigma}/{200\rm~ km~ s^{-1}} \right)^{4.02}M_{\odot}$,
where $\sigma$ is stellar velocity dispersion (Tremaine et al. 2002) if $\sigma$ is known.
We use $\sigma={\rm FWHM_{\rm [O III]}}/2.35$ (Nelson \& Whittle 
1996; see also Boroson 2003; Shields et al. 2003) to estimate the black hole mass according to 
the $M_{\rm BH}-\sigma$ relation.   
This relation is based on evidence that the \oiii line width in AGNs is dominated by the gravitational
potential on the scale of the host galaxy bulge (Nelson \& Whittle 1996; Nelson 2000).
However, it has been shown in SDSS sample that $\sigma={\rm FWHM_{\rm [O III]}}/2.35$ overestimates
the stellar velocity dispersion by a factor of 1.34 and hence the black hole masses (Greene \& Ho 2005). 
We use the corrected $\sigma_*=\sigma/1.34$ to estimate the black hole masses.
Totally we get the black hole masses of 44 NLS1s and 30 non-HBLR S2s, respectively. 
We also estimate $M_{\rm BH}$ of 39 NLS1s with \oiii line width for comparing to those of non-HBLR S2s
with the same method.

Fig. 2 shows the distributions of the black hole masses in NLS1s and non-HBL S2s. We find that the black hole 
masses of the NLS1s distribute in the range of $10^{6.0} - 10^{8.0} M_{\odot}$ with mean values of 
$10^{6.53\pm 0.06}\sunm$ and $10^{6.73\pm 0.11}\sunm$ from H$\beta$ and \oiii width, respectively.
The non-HBLR S2s are in the range of $10^{6.0} - 10^{8.0} M_{\odot}$with a mean value of $10^{6.51\pm 0.19}\sunm$. 
Interestingly, the two distributions are very similar. The black hole masses span the same ranges
and peak at the same values in the two kinds of objects. We use ASURV (Feigelson \& Nelson 1985)  
to show the differences between the two 
distributions within the uncertainties. The H$\beta-$based mass distribution is same with non-HBLR S2s 
with a probability of $p_{\rm null}=55.9\%$\footnote{The probability $p_{\rm null}$ is for the null hypothesis that the
two distributions are drawn at random from the same parent population.}, 
moreover we find that the {\sc [O iii]}$-$determined black hole mass distribution 
is same with  $p_{\rm null}=44.8\%$.   

It is well known that NLS1s usually have less massive black holes, but it is first time to know
that non-HBLR S2s also contain less massive black holes. 
This result also implies that the black hole masses are statistically same in the two kinds of objects.
This is the first piece of basic evidence for that non-HBLR S2s and NLS1s are intrinsically same
populations.  

\figurenum{2}
\centerline{\includegraphics[angle=-90,width=8.0cm]{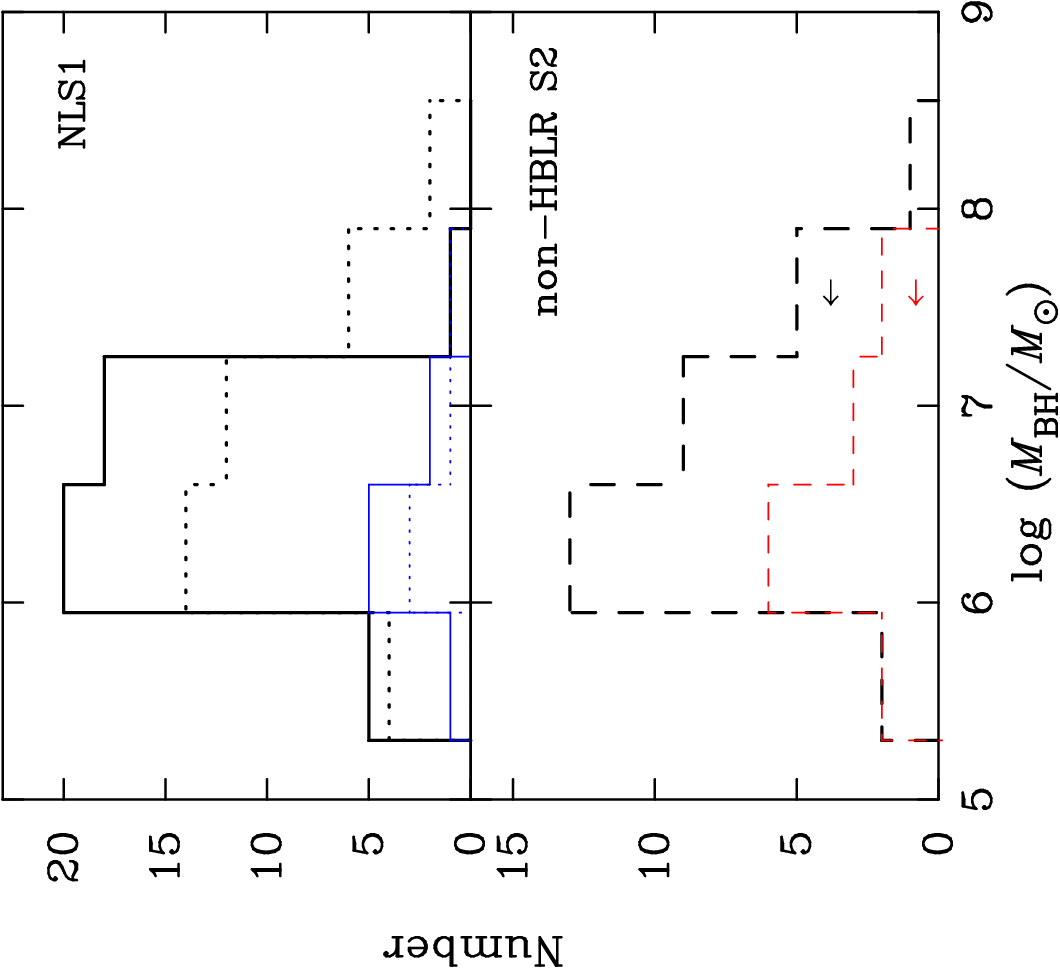}}
\figcaption{The plot of black hole mass distributions. The thick solid/dotted lines indicate that 
BH mass is estimated from the empirical reverberation mapping method and FWHM(\oiii)/$\sigma$, respectively, 
the dashed line for non-HBLR S2s for the present sample. The number of the arrows in each bins is the number of the 
upper/lower limit 
sources.  The thin red-dashed and blue-solid lines represent 12 $\mu$m sample (Rush et al. 1993) and the same in the
subsequent figures.}
\label{fig2} 
\vglue 0.3cm

Why the broad emission lines (BELs) are absent in the non-HBLR S2s? 
Nicastro (2000) suggested that the width of BELs could correspond to the Keplerian velocity of 
an accretion disk at the radius the broad emission line clouds origin. 
The model predicts that for accretion rate $\dot m < 0.2 $ (sub-Eddington 
regime), the FWHMs are quite broad ($> 4000 \rm ~km ~s^{-1}$), while for
$\dot m = 0.2-3$ (from sub-Eddington to moderately super-Eddington regime, 
the corresponding  FWHMs span an interval $\approx 1000-4000 \rm~km~ s^{-1}$. As we show in next section, the
Eddington ratio is quite high in non-HBLS2s.

After determining the black hole masses, we can predict the viral width of the emission line if the distance 
of emitting clouds is known. For virialized clouds emitting H$\beta$ line in the black hole potential well, 
the FWHM can be simply given by (Laor et al. 1997)
\begin{equation}
\Delta v\approx 675 \left(\frac{\mbh}{10^{6.5}\sunm}\right)^{1/2}
                     \left(\frac{L_{\rm bol}}{10^{45}{\rm erg~s^{-1}}}\right)^{-1/2}~~~~{\rm km~s^{-1}},
\end{equation}
where $L_{\rm bol}$ is the bolometric luminosity from the black hole accretion disk and we use the 
relation $R_{\rm BLR}=3.2L_{\rm 45}^{1/2}$pc expected from the photoionization theory. As shown in eq. (2),
the typical FWHM will be of $675~\kms$ for a non-HBLR S2s with
the mean mass of the black holes $\langle \mbh\rangle=10^{6.5}\sunm$ and  
$\langle L_{\rm bol}\rangle=10^{45}$\ergs (see next section).
This means that non-HBLR S2s most likely have a "narrower" broad line region as NLS1s.
This result also strongly implies that non-HBLR S2s are intrinsically same with NLS1s, which
have  less massive black holes.

If the black hole masses are similar in the two kinds of the objects, the bulges of their host galaxies
should be also similar according to the correlation between the bulge and the black hole
(Magorrian et al. 1998;  H\"aring \& Rix 2004). We give the morphologies and bulge magnitudes of their 
host galaxies in Table 2 and 3. Fig. 3. shows the plot of the morphology 
distribution according to Hubble classifications. K-S test shows that the two distributions are same at 
a significance of $23.4\%$.

\figurenum{3}
\centerline{\includegraphics[angle=-90,width=8.0cm]{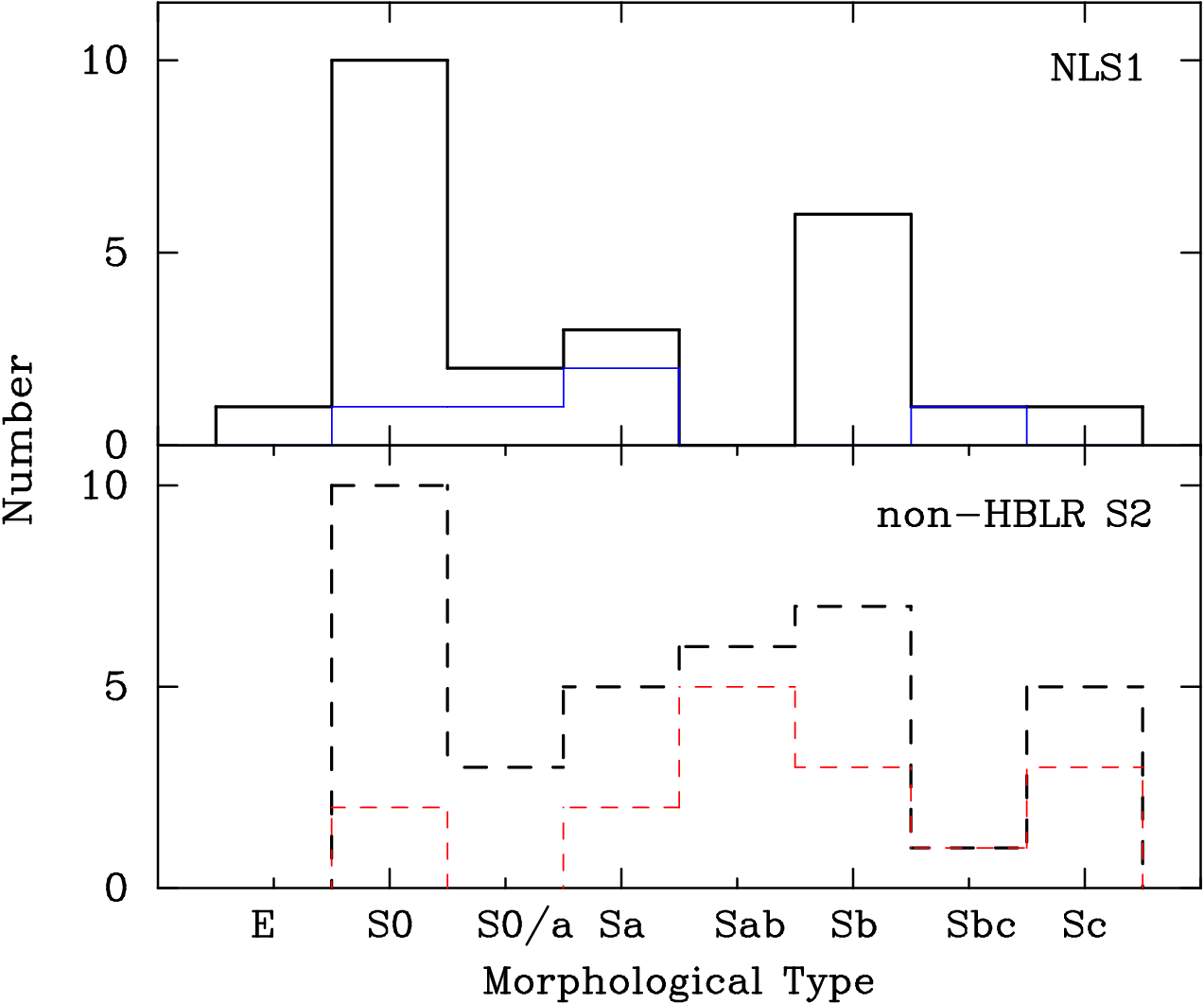}}
\figcaption{The distributions of the morphologies of the host galaxies of non-HBLRS 2s and NLS1s.
The solid line: NLS1s, the dashed line: non-HBLR S2s.}
\label{fig3}    
\vglue 0.3cm 

To further explore the black hole mass distribution, 
we get $B-$band magnitudes of bulges in our sample based on the conversion from the galaxy total
magnitude (Simien \& de Vaucouleurs 1986)
\begin{equation}
M_{\rm bulge}=M_{\rm tot}-0.324\tau+0.054\tau^2-0.0047\tau^3
\end{equation}
where $\tau=T+5$ and $T$ is the Hubble stage of the galaxy. Figure 4 shows distributions of the bulge magnitudes.
K-S test shows that the similarity of the bugle magnitudes is at a level of 26.4\%. This strongly 
indicates that the black hole masses should be similar among the objects of non-HBLR S2 and NLS1s
if the $\mbh-M_{\rm bulge}$ relation works (H\"aring \& Rix 2004).
The similarities in host morphology and bulge magnitudes
lend additional support to the idea that non-HBLR S2s and NLS1s have the same black hole mass.

\figurenum{4}
\centerline{\includegraphics[angle=-90,width=8.0cm]{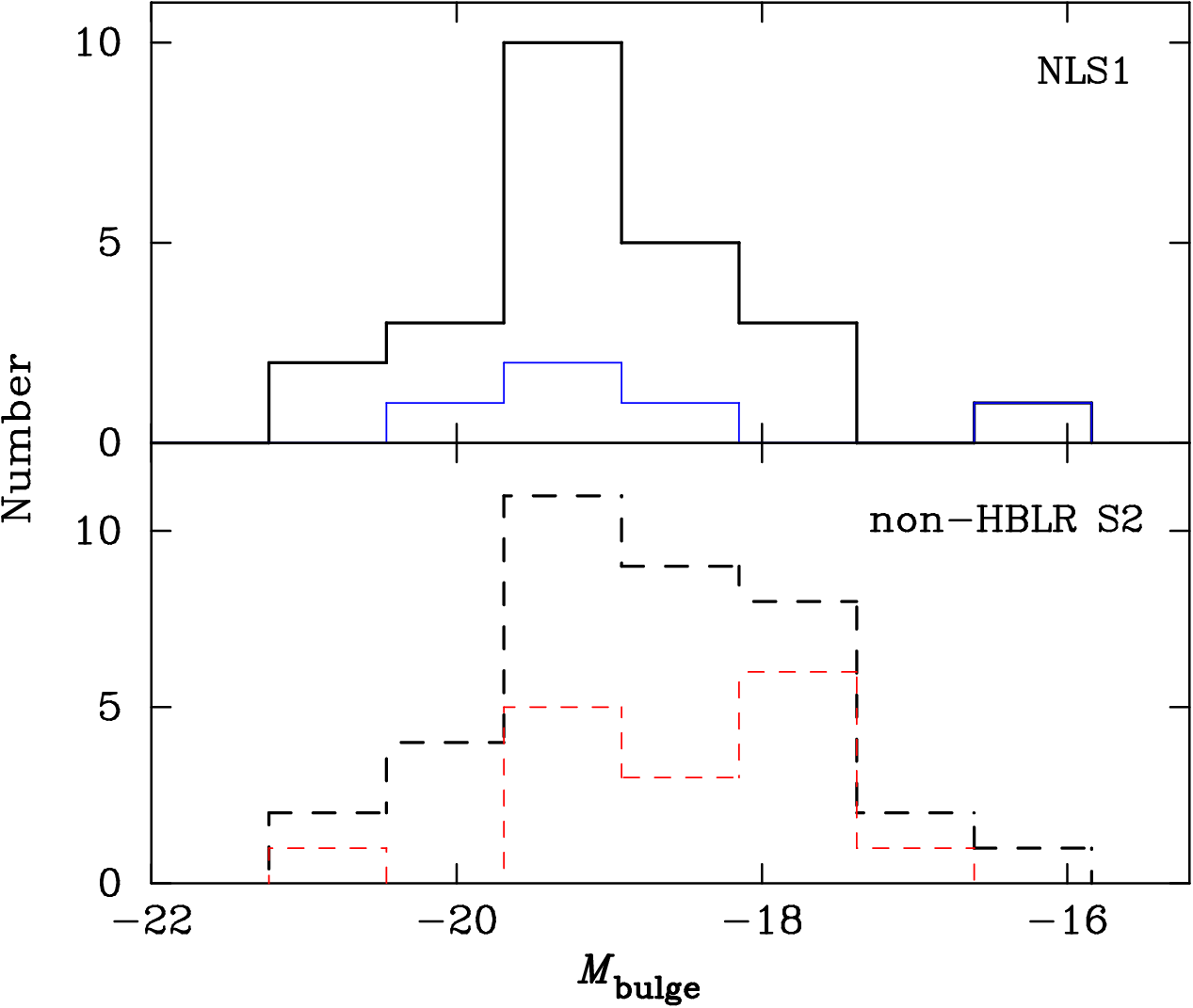}}
\figcaption{The distributions of bulge $B-$band magnitudes of non-HBLRS 2s and NLS1s.
The solid line: NLS1s, the dashed line: non-HBLR S2s.}
\label{fig4}    
\vglue 0.3cm 

There are some uncertainties of the above estimations of the black hole masses. 
For the NLS1s, the empirical reverberation relation gives an uncertainty of  0.5 dex 
(Wang \& Lu 2001). For the non-HBLR S2s, the $M_{\rm BH} - \sigma$ relation has an intrinsic 
dispersion in $M_{\rm BH}$ that is about $\sim 0.3$dex (Tremaine et al. 2002).
The \oiii width can predict the black hole mass to a factor of 5 (Boroson 2003). For the
present non-HBLR S2 sample, we take the uncertainties to be 0.7dex.

We have noted that there are some differences in the mass distributions between two methods and two 
classes of AGNs, especially there is a high mass tail for the \oiii method. This tail is not consistent 
with H$\beta-$method. The possible reasons for these are due to: 1) the \oiii
may compose of complicate components contributed from outflows, which leads to larger uncertainties of the 
black hole mass; 2) the sample of non-HBLR S2s may mix with some of unabsorbed non-HBLR S2s 
since some of them have no X-ray observations. Future work should be improved from the above ways.
  
\subsection{{\rm \oiii}$\lambda 5007$ Line Emission}
It is not realistic to estimate the accretion rates of the black holes in non-HBLR S2s according to
optical and UV continuum since the 
central engines in non-HBLR S2s may be highly obscured. In the UM, the torus is located between 
the broad  and narrow line regions. The \oiii emission is on scale much larger than the torus 
and  would not be affected by the viewing angle. On the other hand, the isotropic \oiii 
luminosity is a good indicator of the ionizing luminosity tightly related with the accretion 
luminosities (Heckman et al. 2004). However as shown by Maiolino \& Rieke (1995), it is not a
completely isotropic indicator for the host galaxy disk might obscure part of the NLR.
The \oiii  luminosity could be taken as an indicator of the nuclear activity only after correction 
for extinction (Maiolino et al. 1998; Bassani et al. 1999; Gu \& Huang 2002).  
The fluxes are corrected for the extinction in non-HBLR S2s and NLS1s by the relation (Bassani et 
al. 1999)
\begin{equation}
F^{\rm cor}_{\rm [O~III]}=F^{\rm obs}_{\rm [O~III]}\left[\frac
{(\rm H \alpha /\rm H \beta)_{obs}}{(\rm H \alpha /\rm H \beta)_{0}}\right]
^{2.94},
\end{equation} 
assuming an intrinsic Balmer decrement $(\rm H\alpha /\rm H \beta)_{0}=2.8$. The corrected luminosities 
are given in Table 2 and 3. The corrections for NLS1s are not straightforward since we have to separate the 
narrow component from the spectrum. This is not an easy job for NLS1s. A detailed treatment for this can be 
found in Dietrich et al. (2005), who investigated the narrow component of
${\rm H\alpha/H\beta}$ for 12 NLS1s. 
They found that the observed ${\rm H\alpha/H\beta}$ is in a range of 2.64 $\sim$ 7.86 with the mean value of 
$\langle {\rm H\alpha/H\beta}\rangle_{\rm obs}=4.88\pm 0.51$. We do not fit the narrow H$\alpha$ and H$\beta$
lines for each objects in detail for this correction, 
but we use the averaged value of $\langle {\rm H\alpha/H\beta}\rangle$ for the entire NLS1s sample.

Fig. 5 shows the \oiii luminosity distributions of the two samples.
The average \oiii luminosities  of the non-HBLR S2s and NLS1s are 
$\langle \log L_{[\rm O~III]}\rangle=41.51\pm0.13$ and 
$\langle \log L_{[\rm O~III]}\rangle=41.60\pm0.09$, respectively. 
The K-S test shows that the luminosities  of \oiii between
these two classes are quite similar ($p_{\rm null} = 49.7\%$).
This result strongly supports the ionizing luminosities indicated by \oiii are same between NLS1s and 
non-HBLR S2s.

\figurenum{5}
\centerline{\includegraphics[angle=-90,width=8.0cm]{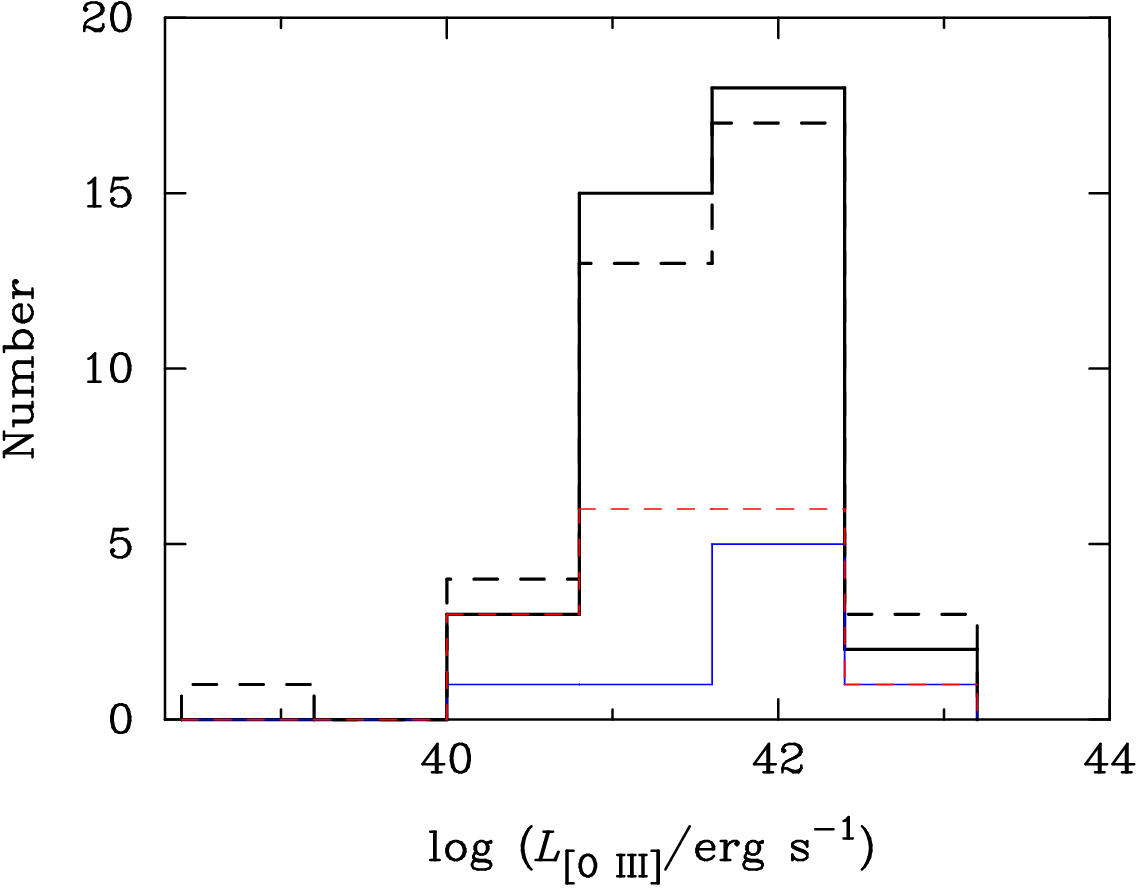}}
\figcaption{The plot of \oiii  $\lambda 5007$ emission line luminosity. The 
solid line: NLS1s, the dashed line: non-HBLR S2s. NGC 5128 is the faintest object of \oiii luminosity
($L_{\rm [O~III]}=10^{38.2}$\ergs).}
\label{fig5}    
\vglue 0.3cm 

Since the \oiii luminosity can be used to probe the ionizing luminosity, it then can be applied to  
estimate the bolometric luminosity of the central engines. 
Heckman et al. (2004) suggested a relation between the \oiii and bolometric luminosities as 
$L_{\rm Bol}\approx 3500L_{\rm [O~III]}$ for Seyfert galaxies. We then have the Eddington ratio through 
\begin{equation}
\cale=\frac{L_{\rm Bol}}{L_{\rm Edd}}
     =1.4\left(\frac{L_{\rm [O~III]}}{3\times 10^{41}{\rm erg~s^{-1}}}\right)
          \left(\frac{\mbh}{10^{6.5}\sunm}\right)^{-1},
\end{equation}
where the Eddington luminosity is given by $L_{\rm Edd}=1.4\times 10^{38}(\mbh/M_{\odot}) \rm erg~s^{-1}$.
We find the typical Eddington ratio is $\cale\sim 1$ in the present sample.

Fig. 6 shows the Eddington ratio distribution of the present sample, implying that most black holes have 
an accretion rate close to or super Eddington limit. Most likely slim disks work in non-HBLR S2s like in NLS1s.
We obtain the averaged values of the Eddington ratios for the NLS1s and non-HBLR S2s from
\begin{equation}
L_{\rm Bol}\approx 9L_{5100}~~~(\rm for~ NLS1s~ only),
\end{equation}
and
\begin{equation}
L_{\rm Bol}=3500L_{\rm [O~III]}~~~({\rm for~ NLS1s~ and~ non-HBLR~ S2s)}.
\end{equation}
For NLS1s, we have $\langle\log\cale\rangle=-0.16\pm 0.05$ from (6) and $\langle\log\cale\rangle =0.19\pm 0.14$
from (7). For non-HBLR S2s, we have
$\langle\log\cale\rangle=0.23\pm 0.14$. We find the similarities at probabilities of $p_{\rm null}=1.6\%$
for $L_{\rm Bol}$ from (6) and (7) and $p_{\rm null}=95.4\%$ from (7) only.  This similarity
implies that similar accretion disks are powering the non-HBLR S2s and NLS1s.
Additionally if the dependence of \oiii luminosity on the Eddington ratio
found in quasars (Baskin \& Laor 2005) applies to the present sample, it suggests that they have similar Eddington 
ratios. Interestingly, Dewangan \& Griffiths (2005) presented evidence of three non-HBLR S2s 
(NGC 5506, NGC 7314 and NGC 7582)\footnote{NGC 5506 (Tran 2003) and NGC 7314 ( Lumsden et al. 2004) are given as 
HBLR S2s, but they are listed as non-HBLR Seyfert 2 galaxies in Dewangen \& Griffiths (2005). We did not include
the two objects in our sample.} being the counterparts of NLS1s based on extremely rapid X-ray variability 
and steep $2 - 12$keV spectrum.
The mirror might reflect the soft X-rays to the observers, showing variability similar to NLS1s.

\figurenum{6}
\centerline{\includegraphics[angle=-90,width=8.0cm]{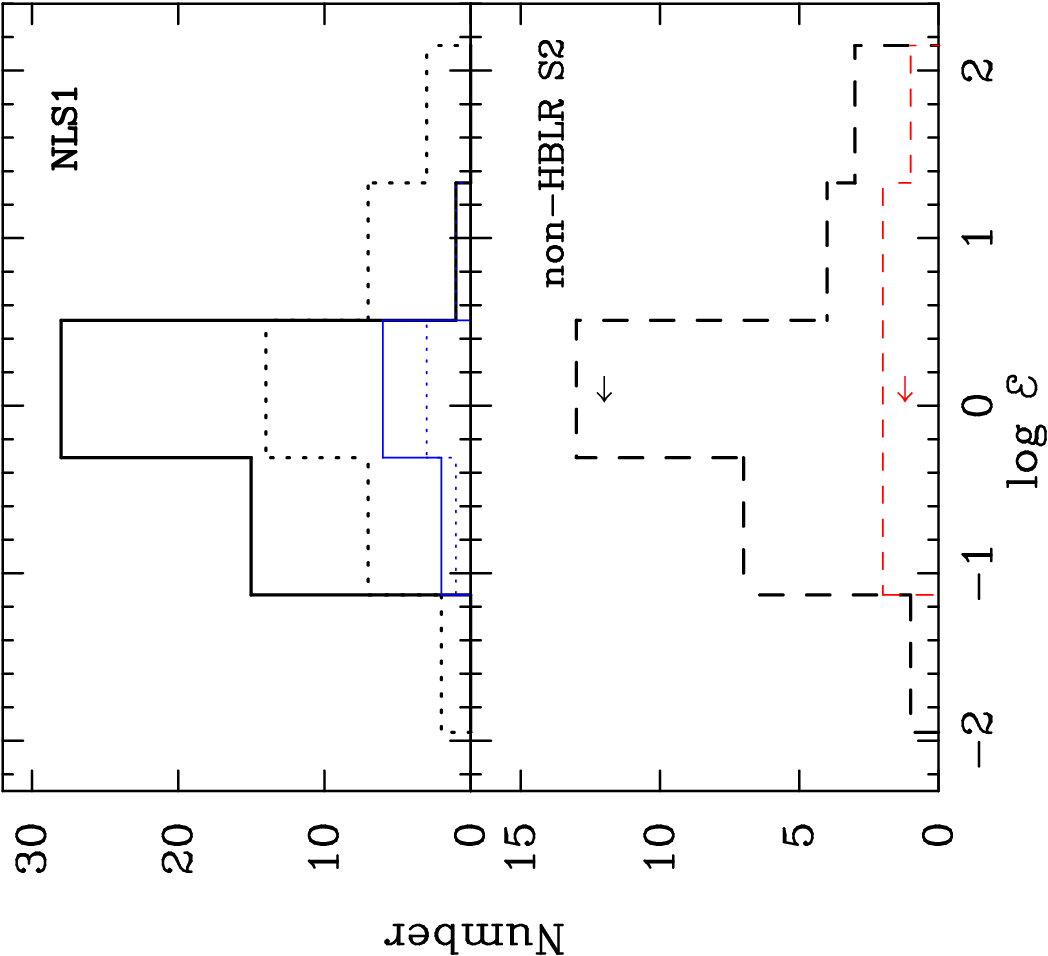}}
\figcaption{The plot of the Eddington ratio distribution in the present sample. The Eddington ratio is 
estimated from \oiii luminosity for both NLS1s and non-HBLR S2s. We also estimate bolometric luminosity from
$L_{\rm Bol}=9L_{\lambda 5100}$. The solid/dotted lines are for NLS1s and the dashed line for non-HBLR S2s.}
\label{fig6}    
\vglue 0.3cm 

Though the similarities of \oiii luminosities, we have to point out 
that there still several potential differences in NLR. There are uncertainties of the same distributions of \oiii 
luminosities in several ways: 1) the NLRs might be not same exactly, such 
as different density, different spectral energy distribution of the ionizing sources, different covering 
factors and different influence of outflow from the center, even the different shocks if it powers partially 
the \oiii luminosity; 2) \oiii luminosity still has a
significant component from the obscured nucleus (Hes et al. 1993; but see Simpson 1998 and
Kuraszkiewicz et al. 2000). However, there are indeed some similarities of NLR in Seyfert galaxies.
The strong correlation between the NLR size ($R_{\rm NLR}$) and the \oiii luminosity (Bennert et al. 
2002; Schmitt et al. 2003; Netzer et al. 2004) indicates that their NLR sizes are similar
in term of the similarity of \oiii luminosities in non-HBLR S2s and NLS1s. Dopita et al. (2002) use 
radiation pressure dominated 
photoionization models to show that efficient \oiii emission comes from regions with $Un_e\approx 1$, where 
the ionization parameter $U=L_{\rm ion}/4\pi cn_eR_{\rm NLR}^2$, $L_{\rm ion}$ is the ionizing luminosity
and $n_e$ is the number density of the electrons. It is expected that the ionizing luminosities are
similar in this way. Moreover, Dietrich et al. (2005) found that NLS1s overlap with 
other Seyfert galaxies in the diagnostic NLR of Veilleux \& Osterbrock (1987). This implies that the NLR  
should be similar to each other. We thus draw a conclusion that $L_{\rm ion}$ is similar in
non-HBLR S2s and NLS1s as shown by Fig. 5, at least, the \oiii luminosity
is a robust prober to examine the NLR as the first order approximation. 
 However we stress here that the present results are produced by the 
first order approximation of the apparent similarity in the \oiii luminosity distribution between NLS1s and 
non-HBLR S2s. 

The two subsections above show strong evidence for the intrinsically same properties of central 
engines ($\mbh$ and $\dot{M}$) in non-HBLR S2s and NLS1s. The two parameters are controlling many aspects
of AGNs, such as metallicity (Shemmer et al. 2005), C {\sc iv} equivalent width and profile
(Baskin \& laor 2005) and the hot corona of the disk (Wang et al. 2004). Here we provide further evidence 
for that their nuclei are intrinsically same between non-HBLR S2s and NLS1s, but the non-HBLR S2s are 
viewed at high inclination angle. The following sections are devoted to show more evidence to support this 
unification scheme.

\subsection{Orientation of Torus}
Though the dusty torus plays a key role in the unification scheme, its geometry remains uncertain.
Geometry of the torus in AGNs may be complicate, even it is likely clumpy (Krolik \& Begelman 1988,
Elizture 2005). As the temperature  decreases  with the radius increase, the 
torus gives longer wavelength radiation from the inner to the outer regions. However, the torus may have 
significant opacity so that only the long wavelength ($\lambda > 50 \mu$m) can escape 
directly (HLB97). Detailed calculations of IR emissions from torus
predict that near/mid infrared radiation from such a 
torus is anisotropic since the reprocessed emission from the inner region of the torus is optically thick 
(Pier \& Krolik 1992; Granato \& Danese 1994). Therefore the observed near/mid infrared emission from the
torus is expected to be an indicator of its orientation (HLB97). 

The FIR emission from Seyfert galaxies may be composed of two components: reprocessed by the torus and the
star formation. We first compare their FIR emissions as shown in Fig. 7. Using ASURV, we get the average of 
the logarithm value $\langle \log L_{60\mu}\rangle=43.76\pm 0.11$ and 
$\langle \log L_{100\mu}\rangle=43.72\pm 0.11$ for non-HBLS2s whereas
$\langle \log L_{60\mu}\rangle=43.66\pm 0.08$ and 
$\langle \log L_{100\mu}\rangle=43.63\pm 0.09$ for NLS1s. The distributions of $60\mu$ and $100\mu$
are similar to each other at a level of $p_{\rm null}=46.2\%$ and 39.4\%, respectively. 
The similarity in FIR emissions show that the two kinds of objects are virtually the same at FIR, or
the emissions from the torus are contaminated by star formation (Alonso-Herrero et al. 2001, Ruiz et al. 2001)
at the same level in NLS1s and non-HBLR S2s.

We define a flux ratio of 
\begin{equation}
{\cal O}=\frac{f_{60\mu{\rm m}}}{f_{25\mu{\rm m}}}, 
\end{equation}
as an orientation indicator of tori (HLB97). The larger ${\cal O}$, the higher inclination to observers.
Fig. 8 shows the distributions of the ratio ${\cal O}$ in the present sample.
The distributions of the ratio ${\cal O}$ show great differences between NLS1s and non-HBLR 
S2s. Using ASURV, we get the mean values $\langle {\cal O}\rangle=2.68 \pm 0.27$ and 
$\langle {\cal O}\rangle=4.99 \pm 0.41$ for NLS1s and non-HBLR S2s, respectively. 
%The K-S test shows the two distributions are 
The  two distributions are same at a probability of $p_{\rm null}= 10^{-4}$.

\figurenum{7}
\centerline{\includegraphics[angle=-90,width=8.0cm]{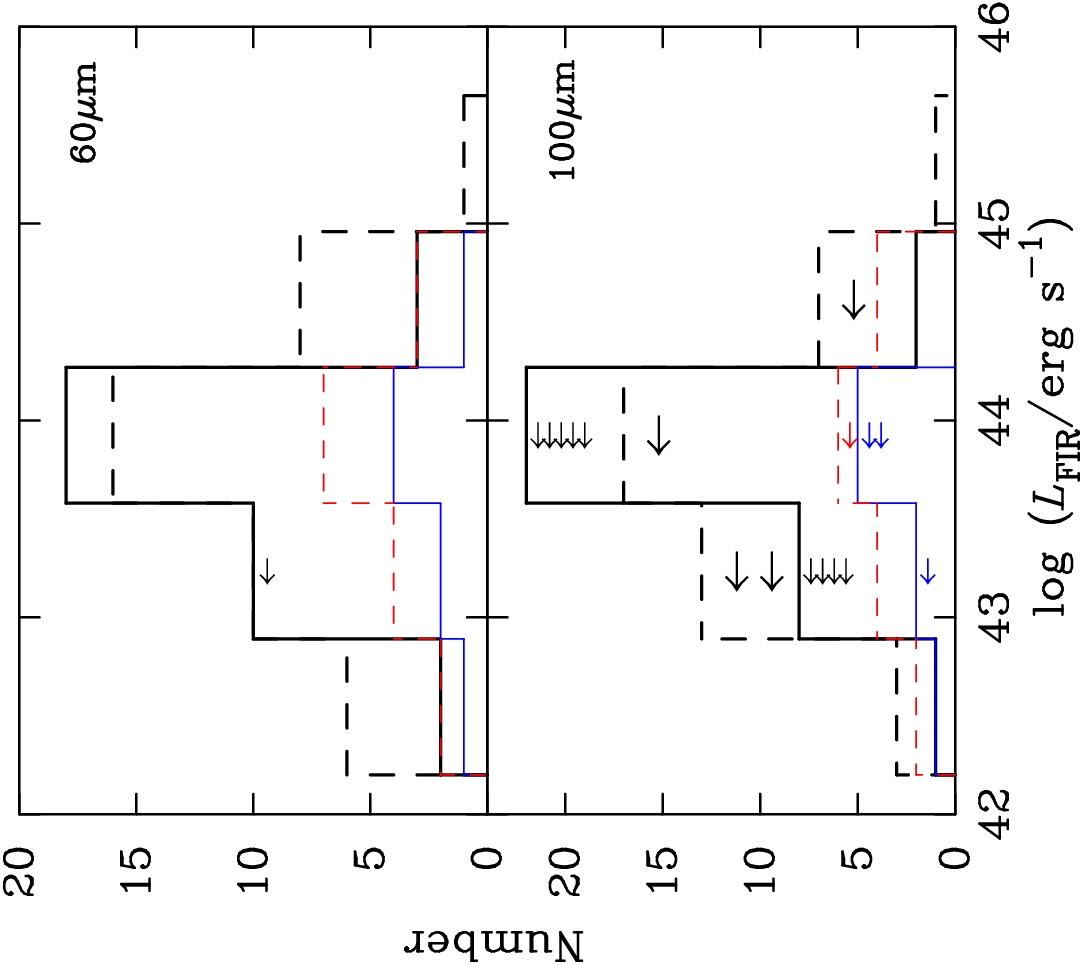}}
\figcaption{The plot of far-infrared luminosity. The solid line: NLS1s, the 
dashed line: non-HBLR S2s. }
\label{fig7} 
\vglue 0.5cm
\figurenum{8}
\centerline{\includegraphics[angle=-90,width=8.0cm]{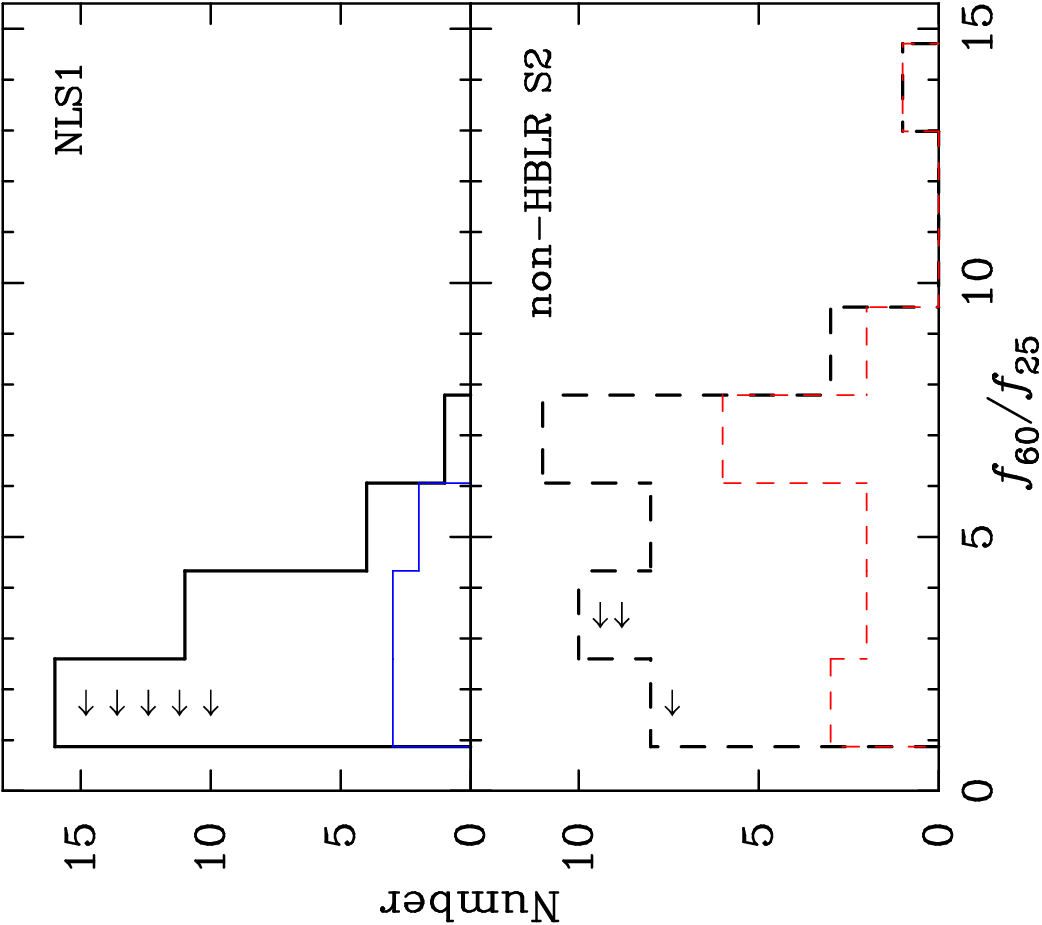}}
\figcaption{The plot of infrared flux ratio ${\cal O}$. The solid line: NLS1s, the dashed line: non-HBLR S2s. }
\label{fig8} 
\vglue 0.3cm

As we have shown, the distributions of the far-infrared luminosities are indistinguishable in the two 
samples. Though we do not separate the reprocessed components from star formation, the similar distributions 
allow us to use it as a normalization factor for orientation of the torus in the present sample.
The different ${\cal O}-$distributions indicate that orientations are different,
showing that the non-HBLR S2s tend to be viewed at high inclination angles whereas the NLS1s at
low inclination angles. 

Whether this parameter represents the orientation of the dusty torus still remains as a debate in the 
literatures (see Alexander 2001; Gu et al. 2001).  We note that all these authors compared  HBLR S2s 
and non-HBLR S2s, i.e. type 2, and these two classes tend to have large view angle. They did not 
compare the ratio between type 1 and type 2. Tran (2003) compared ${\cal O}-$distributions among broad
line Seyfert 1s, HBLR S2s and non-HBLR S2s. He found that non-HBLR S2s are quite different from 
HBLS2s and Seyfert 1s are similar to HBLR S2s. He suggested that this ratio is not significantly
affected by orientation, namely, it is not good indicator of the torus.
We have to point out that Tran's sample is mixed with NLS1s (6 objects) and unabsorbed Seyfert 2s 
(at least 3 objects: NGC 3660, NGC 4501 and NGC 5929) and his sample is much smaller than ours in the 
present sample.  The parameter ${\cal O}$ stands for torus orientation in the present sample.  

\subsection{Radio Emission}
Radio emission from Seyfert galactic core (within $\sim 100$pc) can also penetrate the torus and has
potential isotropic property. Early detections of radio emission show evidently stronger radiation from 
Seyfert 2 galaxies than in Seyfert 1s, however it has been realized that these differences are likely due 
to selection bias as shown in later studies (Giuricin et al. 1990, Thean et al. 2001, Ulvestad \& Ho 2001).
Palomar Seyfert galaxies show they obey the same radio-forbidden line relations (Ulvestad \& Ho 2001).

\figurenum{9}
\centerline{\includegraphics[angle=-90,width=8.0cm]{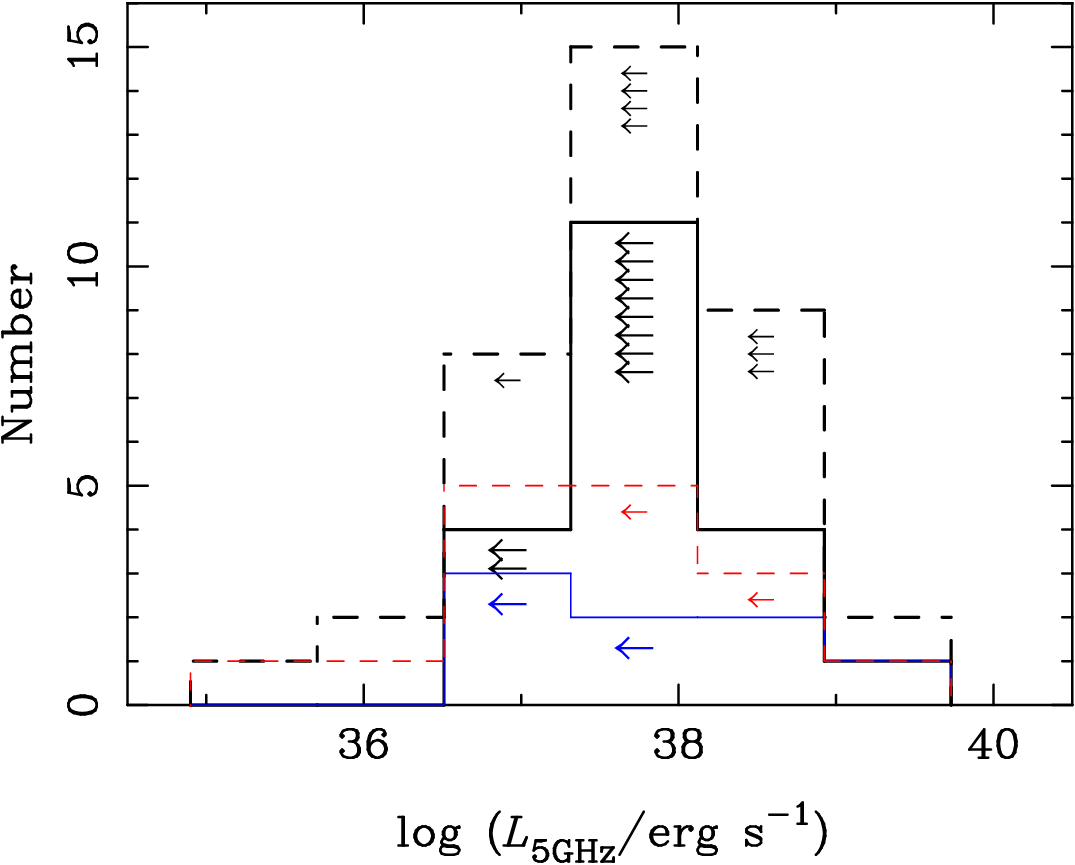}}
\figcaption{The plot of radio (5.0GHz) emission luminosity. The solid line: 
NLS1s, the dashed line: non-HBLR S2s. }
\label{fig9} 
\vglue 0.3cm

Fig. 9 shows the distributions of radio luminosities in non-HBLS2s and NLS1s.
We find that the two overlap nicely and have a similar peak luminosity. This similarity of radio luminosity 
distributions show the radio activities are similar. In our sample, the average values of radio luminosities 
are $\langle \log L_{5\rm GHz}\rangle= 37.45\pm 0.20$ and 
$\langle \log L_{5\rm GHz}\rangle= 37.49\pm 0.16$ for  NLS1s and non-HBLR S2s, respectively. 
Using ASURV,  we get the probability of the two samples to be extracted from 
the same parent population $p_{\rm null}=75.9\%$. 
The strength of the radio cores shows a similar level of nuclear  activity and may 
indicate the masses of supermassive black holes (Thean et al. 2001; Franceschini, 
Vercellone \& Fabian 1998).
The similarity of radio luminosities of NLS1s and non-HBLR S2s is consistent 
with the prediction of UM and the above results of $M_{\rm BH}$ and accretion rates.

Ho \& Ulvestad (2001) and Ulvestad \& Ho (2001) investigated radio emissions from 
an optically selected sample of Palomar Seyfert galaxies. They
find there is a very strong correlation between the radio and the \oiii luminosity
as $L_{\rm 6cm}\propto L_{\rm [OIII]}^{0.8}$.
Fig. 10 shows the $L_{\rm 5 GHz}-L_{\rm [O III]}$  correlation  for both samples given in Table 4. 
We find that these correlations agree
with that found in Ulvestad \& Ho (2001) for Polomar Seyfert galaxies. This result shows that
non-HBLR S2s and NLS1s obey the same radio $-$ \oiii relation, implying that the radio emission
is powered by the same mechanism in non-HBLR S2s and NLS1s, even as in the normal Seyfert galaxies
(Ulvestad \& Ho 2001).

\figurenum{10}
\centerline{\includegraphics[angle=-90,width=8.0cm]{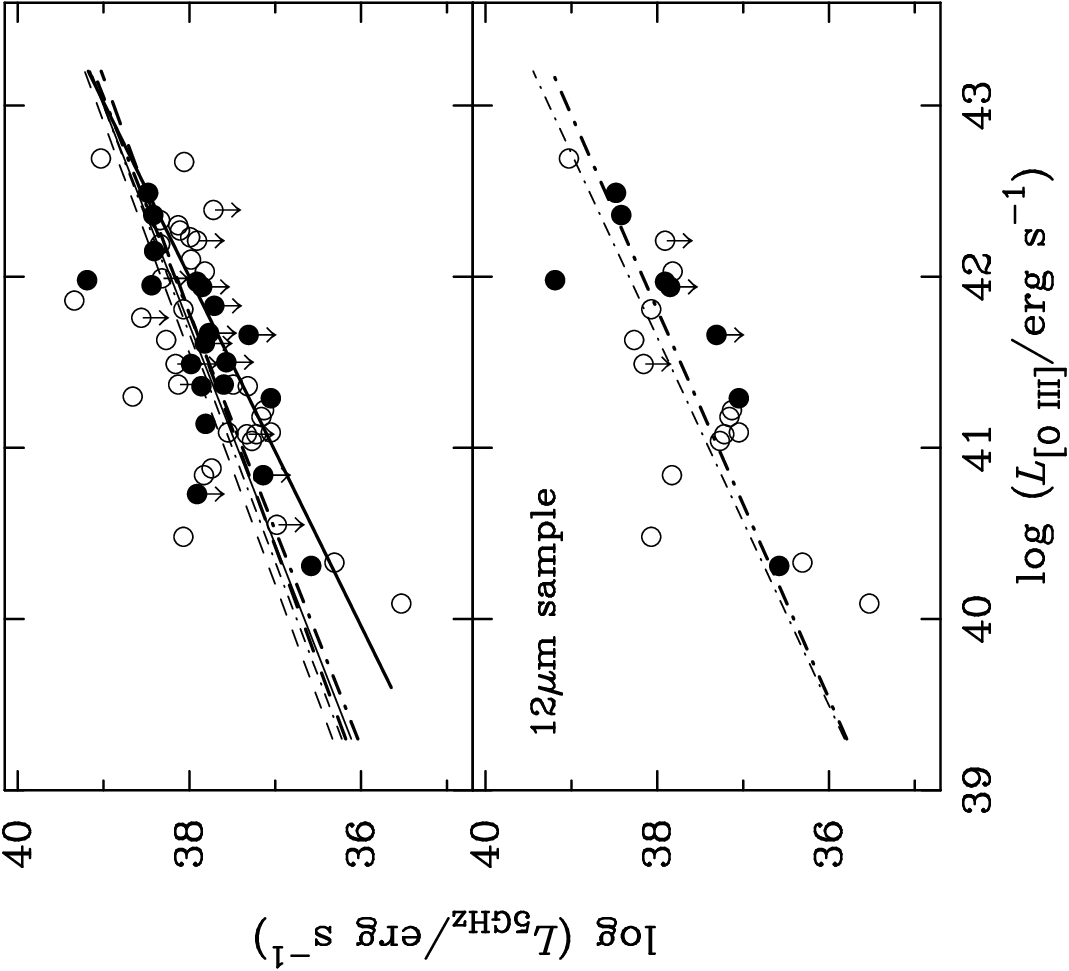}}
\figcaption{{\em Upper}: The plot of the correlation between the radio (5.0GHz) emission 
and \oiii luminosities. Filled circles represent NLS1s whereas the open the non-HBLR S2s.
The solid, dashed and dashed-dotted lines (thin)  are for NLS1s, non-HBLR S2s and the entire 
sample (excluding upper limit sources), respectively. Thick lines are for the entire sources.
{\em Lower}: The correlations for 12 micron sample. The dashed-dotted line (thin) exludes
the upper sources whereas the thick for the entire sources.
}
\label{fig10} 
\vglue 0.3cm

\begin{center}
\footnotesize
\centerline{\sc Table 4. The Correlation Analysis}
\begin{tabular}{lcccc}\hline \hline
\multicolumn{5}{c}{ Analysis with censored data}\\ \hline
objects         &     $q$        &     $k$      & $\rho$&    $p$              \\ \hline          
NLS1            &$-3.16\pm9.45$  &$0.98\pm0.23$ &$0.62$ &$6.9\times 10^{-3}$  \\
non-HBLRS2      &$ 7.49\pm6.24$  &$0.73\pm0.15$ &$0.56$ &$1.4\times 10^{-3}$  \\
total           &$ 4.99\pm5.22$  &$0.79\pm0.13$ &$0.59$ &$<1.0\times 10^{-4}$ \\ 
total(12$\mu$m) &$ 1.21\pm6.89$  &$0.88\pm0.17$ &$0.60$ &$ 4.6\times 10^{-3}$ \\ \hline 
       \multicolumn{5}{c}{  Analysis without censored data }               \\ \hline
NLS1            &$5.46\pm6.72$   &$0.78\pm0.16$ &$0.69$ &$2.6\times 10^{-3}$  \\
non-HBLRS2      &$7.25\pm5.55$   &$0.74\pm0.13$ &$0.62$ &$4.0\times 10^{-4}$  \\
total           &$6.75\pm4.32$   &$0.75\pm0.10$ &$0.67$ &$<1.0\times 10^{-4}$ \\
total(12$\mu$m) &$-0.73\pm6.15$  &$0.93\pm0.15$ &$0.74$ &$5.0\times 10^{-4}$  \\ \hline 

\end{tabular}
\parbox{3.3in}
{\baselineskip 9pt
\noindent 
{\sc Note:} 
$\log L_{\rm 5GHz}=k \log L_{[\rm O III]}+q$,  
 $\rho$ is the Spearman's coefficient and $p$ is the null-probability.
}
\end{center}

\normalsize

We have to stress that our discussions on radio emission from the two samples are suffering from Malquist
bias since only roughly half of NLS1s have radio data. The fraction of radio-loud NLS1s is 
extremely low (Greene, Ho \& Ulvestad 2006, Komossa et al. 2006), so the present results should hold 
if we have more data.

We note that there are a couple of radio-loud NLS1s according to their radio loudness, for example
PKS 2004-447 (Oshlack et al., 2001);  PKS 05548-540, RXJ 0134-4258 
(Komossa et al. 2005). There is growing evidence for presence of radio loud NLS1s with radio-loudness
${\cal R}>10$, however the fraction of radio-loud NLS1s is only $5\sim 6\%$ in SDSS sample (Greene, Ho \& 
Ulvestad 2006), $\sim 7\%$ in Catalogue of Quasars and Active Galactic Nuclei (Komossa et al. 2006). This 
low fraction strongly supports that the higher Eddington ratio may suppress the formation of the jet (Ho 2002). 
Such a tendency is also consistent with blazars, namely the jet becomes weaker with increasing
Eddington ratios (Wang, Luo \& Ho 2004). 
Future studies on this subject will help to understand the differences between NLS1s and non-HBLR S2s.

\subsection{Summary} 
Table 5 summaries the statistical properties of the non-HBLS2 and NLS1s in the
present sample. The first column lists the parameters explored in this paper, the second and third give the
average values of the parameters for non-HBLSR2s and NLS1s, respectively, the 4th does the
null probability for the different distributions of the two kinds of the objects and the last provides a note
on the parameter properties.

{\footnotesize
\begin{center}{Table 5. A Summary of the Statistical Properties}
\begin{tabular}{lccll}\hline \hline
Parameters                             &   non-HBLS2s  & NLS1s          & $p_{\rm null}$ &Note\\ \hline
$\langle M_{\rm bulge}\rangle$         &$-18.71\pm0.17$&$-19.05\pm0.20$ & 26.4\%         & isotropic  \\
$\langle \log \mbh\rangle^a$           &$6.51\pm0.19$  &$ 6.53\pm0.06$  & 55.9\%         & isotropic  \\
$\langle \log \mbh\rangle^b$           &$6.51\pm0.19$  &$ 6.73\pm0.11$  & 44.8\%         & isotropic  \\
$\langle \log {\cal E}\rangle^c$       &$0.23\pm0.14$  &$-0.16\pm0.05$  &  1.6\%         & isotropic  \\
$\langle \log {\cal E}\rangle^d$       &$0.23\pm0.14$  &$ 0.19\pm0.14$  & 95.4\%         & isotropic  \\
$\langle f_{60}/f_{25}\rangle$         &$4.99\pm0.41$  &$ 2.68\pm0.27$  &$10^{-4}$       & anisotropic\\
$\langle \log L_{\rm [O~III]}\rangle$  &$41.51\pm0.13$ &$41.60\pm0.09$  & 49.7\%         & isotropic  \\
$\langle \log L_{\rm 5GHz}\rangle$     &$37.49\pm0.16$ &$37.45\pm0.20$  & 75.9\%         & isotropic  \\
$\langle \log L_{\rm 60}\rangle$       &$43.76\pm0.11$ &$43.66\pm0.08$  & 46.2\%         & isotropic  \\ 
$\langle \log L_{\rm 100}\rangle$      &$43.72\pm0.11$ &$43.63\pm0.09$  & 39.4\%         & isotropic  \\ \hline
  
\end{tabular}
\parbox{3.3in}
{\baselineskip 9pt
\noindent
{{\sc Note}: $^a$based on empirical relation of reverberation, $^b$on \oiii width, 
$^c$on eq. (7), $^d$based on 5100\AA~ luminosity. } 
When there are censored data, we use Gehan's generalized Wilconxon 
test (hypergeometric variance) in ASURV.
}
\end{center}
}

\figurenum{11}
\centerline{\includegraphics[angle=-90,width=8.0cm]{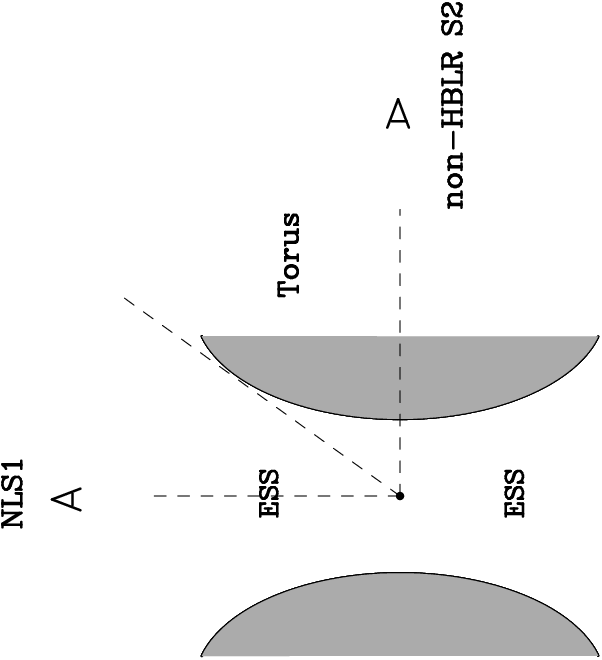}}
\figcaption{The plot of suggested unified scheme. ESS: Electron Scattering Screen. The
location of the ESS and its size remain open (see detail in discussions).}
\label{fig11}   
\vglue 0.3cm

The carton of Fig. 11 shows the unified model of NLS1s and non-HBLS2s. It is well-known that there are 4 classes 
of  objects (we do not include the unabsorbed Seyfert 2 galaxies here): BLS1s, NLS1s, HBLR S2s and non-HBLR S2s. 
This hypothesis alleviates the challenges to the classical unified model.
If the two couples of BLS1 $-$ HBLR S2s and NLS1s $-$ non-HBLR S2s 
are different only in view direction as in the UM,  what are the relations between NLS1s --- BLS1s (Kawaguchi et 
al. 2004) and non-HBLR S2s --- HBLR S2s? Tran (2003) noted that, for powerful narrow-line radio galaxies and radio
weak LINERS, the higher the radio power of the objects the higher the fraction of AGNs found to possess HBLRs. 
The non-HBLR S2s and HBLR S2s may simply be different stages of the evolutionary path. The BLR arose in AGNs 
when the activity reaches above a threshold (Nicastro 2000). The reasonable evolutionary sequence would be  from NLS1s 
to BLS1s and non-HBLR S2s to HBLR S2s. Although the dust torus as well as its orientation accounts for many 
observations, its evolutionary consequence should be another key to understand the UM (Tran 2003, Wang 2004,
Zhou \& Wang 2005, Wang et al. 2005). 
It is obvious that the unification scheme with black hole growth will provide such a connection 
(Wang et al. in preparation).  

\section{Discussions}
\subsection{Limitation of the Present Sample}
Spectropolarimetric observations are time-consuming and tedious, so there is limit number of non-HBLR S2s 
in the literatures. This sets up the criterion of $z\le 0.05$ for the present sample. The redshift-limited 
selection in this paper avoids the luminosity bias. The main goal of the present paper is to show the unification 
of the {\em absorbed}  non-HBLR S2s and NLS1s. However about $40\%$ of the non-HBLR S2s in the present sample do not 
have X-ray observations. This could result in some of heterogeneities of the present sample. If the unabsorbed Seyfert 2 
galaxies has a fraction of $10-30\%$ of Seyfert 2 galaxies (Panessa \& Bassini 2002) and $50\%$ of Seyfert 2 galaxies 
are non-BLR S2s (Tran 2003), it is then expected that there are only about $\sim 3$ non-HBLR S2s are unabsorbed in the
present sample. So we expect that the present sample is complete above a level $93\%$ for the absorbed non-HBLR S2s.
Regarding the heterogeneity of the non-HBLR S2 sample, it may origin from three possible ways. First,  the sample mainly 
contains those with less massive black holes and higher Eddington ratio $\cale\sim 1$.  Actually  these objects occupy
the large parts of the sample since we exclude those without absorption in X-ray band if known. Second, it covers some of
those evolved from the polarized broad line Seyfert 2 galaxies when the black holes tend to exhaust the supplied fuel from 
the dusty torus (Krolik \& Begelman 1988). They thus have more massive black holes and have no absorption. In such a case 
the non-HBLR S2s are real Seyfert 2 since their accretion rates are lower than a critical value 
$\dot{m}_{\rm cr}\approx (1-4)\times 10^{-3}$ so that the broad line region is {\em not} able to form (Nicastro et al. 
2003).  Third, the sample also includes some of those evolved from the broad line Seyfert 1 galaxies on tending to 
depleting the fuel. Such a case is indistinguishable to the second case from absorptions and line width. The present
sample is not pure since some of them have not been detected by X-ray observations. The exact fractions of the second 
and third objects are unknown, 
but we believe that the present sample covers the most of the absorbed non-HBLR S2s.

\subsection{Electron Scattering Screen (ESS)}
The ESS generally plays an important role in the unification scheme. However, in the carton of the unification 
scheme for NLS1s and non-HBLR S2s, we do not plot it since its location is mostly uncertain. 
Three kinds of the problems are to be clarified. First we do not know whether it is always there. This deals 
with the formation of the ESS. Second we do not know the location of the ESS, especially Miller 
\& Goodrich (1991) suggest that it does not
hold a static hydro equilibrium. Third the ratio of ionized gas to dust, scattering depth and geometric size 
are uncertain. We only briefly discuss these problems here and leave them in a future paper.

Two possible ways to form the ESS are: 1) the hot gas evaporated by the central ionizing
source (or originates from the transition region\footnote{How to switch on the geometrically thin disk from 
the thick torus remains an open question. The undergoing physical process must be complicate in this region.} between 
the geometrically thin disk and thick torus) at the inner edge of the dusty torus ; 
2) disk winds, namely a moving ESS.  Accordingly the location of the
ESS will be different. The spatially resolved several nearby galaxies, such 
as in NGC 1068 (Capetti et al. 1995, see a brief review of Kishimoto et al. 
2004), showing a dimension of the ESS $\sim 100$pc. Taniguchi et al. (1999) suggest that the non-HBLS2s are
caused by the more compact ESS, which is still obscured by the dust torus. In such a case the ESS may be
part of the most inner edge of the  evaporated dusty torus. If so, the polarized Seyfert 2 
galaxies will be the objects viewed at an angle intermediate between Seyfert 1 and non-HBLR S2s. This will
definitely lead to a very small fraction of the HBLR S2s. This does not seem happen. A wider range of the 
location is preferred for that the properties of Seyfert 2 galaxies are quite scatter. 
Smith et al. (2004) suggest a more complicate
geometry of the ESS, in which equatorial scattering and polar scattering regions are responsible for the
scattering the obscured broad line to observers. The soft X-ray 
from non-HBLR S2s may be scattered by the ESS to observers and is expected to detect. 
We stress here that the ESS location plays an important role in the reflection of soft X-ray emissions. 
If its location is too close to the center, the scattered soft X-rays are much weaker than that for the 
case of ESS at higher location. Detail studies of such an effect are necessary. 

The geometry and composition of the reflecting mirror determine degrees of the spectropolarimetric
observations. Polarization due to hot dust grains more strongly depends on their geometry and size (Onaka 1995).
If the gas-dust ratio is $g$, the resultant spectrum of an emission line from the reflection region will be
strongly modified by $g$ (Maiolino et al. 2001). However whether the mirror holds hydro-equilibrium remains
open, most likely depending on the luminosity of the central source. The absence of the polarized broad lines
in non-HBLR S2s is caused by the less massive black holes and higher accretion rates, but the properties
of the ESS is worth studying in future.

\section{Conclusions} 
We assemble two sub-samples of consisting of 44 NLS1s and 42 non-HBLR S2s to examine whether they have similar
or same central engines. Most of NLS1s
in the present sample have a steep soft X-ray spectrum ($\Gamma_{\rm SX}>2$) whereas the non-HBLR S2s have 
strong absorption in X-ray band. We estimate black hole masses and accretion rates of these samples. 
We find that: 1) the two kinds of Seyfert galaxies have same distributions of black hole masses from 
FWHM([O {\sc iii}])
at $p_{\rm null}=44.8\%$ in ASURV test; 2) the \oiii luminosities as an accretion rate indicator 
are similar and the black holes have accretion rates close to or above the Eddington limit; 3) the
$f_{60\mu}/f_{25\mu}$ as orientations of the dusty torus are very different. We suggest that non-HBLR S2s 
are counterpart of NLS1s at edge-on orientation.  This hypothesis is further supported by the comparison 
with other indicators. NLS1s and non-HBLR S2s can be unified based on orientation, but they have
less massive black holes with higher accretion rates different from the broad line Seyfert 1 galaxies and HBLR S2s. 
This hypothesis sets up a scenario of a population of less massive black holes at {\em all} orientations, which
have higher accretion rates and are evolving to broad line Seyfert galaxies.
Future work on unification scheme with the growth of the black 
holes and electron scattering screen are need to be done in future.

\acknowledgements{J.M.W\@. is grateful to L. C. Ho for a large number of useful comments on the early version of
the manuscript, which greatly improve its quality.
S. Komossa is acknowledged for discussions. The authors thank the anonymous referee for helpful comments.
Y.-M. Chen is thanked for checking some data. We thank simulated discussions among people in IHEP AGN group. 
The research is
supported by NSFC through NSFC-10325313, NSFC-10233030 and NSFC-10521001.}

\clearpage

\begin{landscape}

%\begin{deluxetable}{lcccccccccccccccl}

%\tablewidth{0pt}
%\tablecaption{{\sc The Narrow Line Seyfert 1 Sample}
%\label{tbl-2}}
%\tablehead{
%Name&$z$& morphology&$M_{\rm bulge}$ & FWHM &$\log\mbh$&$\log \cal{E}$& FWHM &$\log\mbh$&$\log \cal{E}$&$\log L_{\rm [O~ III]}$&$\log L_{\rm 5GHz}$ &$f_{25}$&$f_{60}$&$f_{100}$&$\Gamma_{\rm SX}$&Ref.\\ 
%    &   &               &  &H$\beta$&  H$\beta$ &    H$\beta$      &\oiii &\oiii      & \oiii            &                      &                    &        &        &         &      &     \\
% (1)&(2)&        (3)    &   (4)  &      (5)  &   (6)            & (7)  &  (8)      &  (9)             &           (10)       &          (11)      &  (12)  &  (13)  &  (14)   & (15) & (16)& (17) }
%                                                  
%\startdata

\begin{center}{\sc \small Table 2. The Narrow Line Seyfert 1 Sample}
\footnotesize
\vglue 0.2cm
\begin{tabular}{lcccccccccccccccl}\hline\hline
Name&$z$& morphology&$M_{\rm bulge}$ & FWHM &$\log\mbh$&$\log \cal{E}$& FWHM &$\log\mbh$&$\log \cal{E}$&$\log L_{\rm [O~ III]}$&$\log L_{\rm 5GHz}$ &$f_{25}$&$f_{60}$&$f_{100}$&$\Gamma_{\rm SX}$&Ref.\\ 
    &   &               &  &H$\beta$&  H$\beta$ &    H$\beta$      &\oiii &\oiii      & \oiii            &                      &                    &        &        &         &      &     \\
 (1)&(2)&        (3)    &   (4)  &      (5)  &   (6)            & (7)  &  (8)      &  (9)             &           (10)       &         
 (11)      &  (12)  &  (13)  &  (14)   & (15) & (16) & (17) \\ \hline

 1340+569                & 0.040 & ...        & $  ...  $ & 1500 &  6.36 &  $-0.48$ & 400 & 7.34 & $-0.85$ & 41.08 &  ...           & ...     &  ...    &  ...    & ...                  &     7,  7,  7             \\ 
 Akn564                  & 0.025 & SBb        & $-19.11 $ &  750 &  6.27 &  $ 0.43$ & 220 & 6.29 & $ 1.26$ & 42.15 & 38.41$^{a}$    & 0.57    & 0.83    &  1.14   & $3.47\pm0.07$        & 4, 30,  1, 14, 45, 13, 30 \\ 
 ESO12-G21$^{\ast}$      & 0.030 & ...        & $  ...  $ & 1000 &  6.70 &  $ 0.32$ & ... & ...  & $  ...$ & 41.94 & $<37.85$$^{b}$ & 0.25    &  1.45   & 2.98    & $3.38\pm0.90$        &    52,     53, 25, 23, 53 \\ 
 Fairall303              & 0.040 &  Ep        & $-18.78 $ & 1450 &  6.40 &  $-0.49$ & 140 & 5.50 & $ 1.59$ & 41.69 &  ...           & 0.11    &  0.28   & $<0.29$ & $1.51   $            &13,  9,  9,  9,     11,  9 \\ 
 HS0328+0528             & 0.046 & ...        & $  ...  $ & 1590 &  6.67 &  $-0.42$ & 220 & 6.29 & $  ...$ &  ...  &  ...           & ...     &  ...    &  ...    & ...                  &    30,  1                 \\ 
 HS1831+5338             & 0.039 & ...        & $  ...  $ & 1555 &  6.74 &  $-0.33$ & 240 & 6.45 & $  ...$ &  ...  &  ...           & ...     &   ...   &  ...    & ...                  &    30,  1                 \\ 
 IC3599                  & 0.021 & Sb? pec    & $-17.49 $ & 1200 &  6.28 &  $-0.29$ & 580 & 7.99 & $-1.57$ & 41.01 &  ...           & $<0.13$ &  0.15   & $<0.40$ & $3.20\pm0.10$        &30, 30, 10, 10,     11, 11 \\ 
 IRAS03450+0055$^{\ast}$ & 0.031 & ...	      & $  ...  $ & 1310 &  6.46 &  $-0.29$ & ... & ...  & $  ...$ & 41.98 & 39.19$^{c}$    & 0.51    &  0.47   & $<3.24$ & ...                  &    30,     15, 27, 13     \\ 
 IRAS04312+4008          & 0.020 & S          & $-18.31 $ &  690 &  5.87 &  $ 0.24$ & 380 & 7.25 & $  ...$ &  ...  &  ...           & 0.56    &  2.68   & 3.86    & $2.82_{0.65}^{0.59}$ &30, 30,  1,         13, 49 \\ 
 IRAS04576+0912          & 0.037 & ...        & $  ...  $ & 1220 &  6.35 &  $-0.26$ & 380 & 7.25 & $  ...$ &  ...  &  ...           & 0.33    &  1.65   & 2.33    & $1.43_{0.44}^{2.33}$ &    30,  1,         13, 49 \\ 
 IRAS04596-2257          & 0.041 & ...        & $  ...  $ & 1500 &  6.81 &  $-0.23$ & ... & ...  & $  ...$ & 42.84 &  ...           & 0.25    &  0.88   & 1.18    & ...                  &    30,             13     \\ 
 IRAS05262+4432          & 0.032 &  S         & $-20.94 $ &  700 &  6.47 &  $ 0.69$ & 365 & 7.18 & $  ...$ &  ...  &  ...           & 0.27    &  1.87   & 3.72    & ...                  &30, 30,  1,         13     \\ 
 IRAS15091-2107$^{\ast}$ & 0.044 & ...        & $  ...  $ & 1480 &  7.01 &  $-0.05$ & ... & ...  & $  ...$ & 42.49 & 38.48$^{b}$    & 0.50    &  1.52   & 1.55    & $2.61\pm1.04$        &    30,     12, 25, 13, 50 \\ 
 IRASF12397+3333         & 0.044 & ...        & $  ...  $ & 1640 &  6.66 &  $-0.48$ & 485 & 7.67 & $-0.17$ & 42.10 &  ...           & 0.19    &  0.34   & $<0.76$ & $2.02$               &     9,  9,  9,     13,  9 \\ 
 KUG1031+398             & 0.042 & compact    & $  ...  $ &  935 &  6.41 &  $ 0.19$ & 315 & 6.92 & $ 0.05$ & 41.57 &  ...           & $<0.17$ &  0.35   & 0.66    & $4.15\pm0.10$        &     1,  1,         13, 30 \\ 
 KUG1618+410             & 0.038 & Spiral     & $-20.71 $ & 1820 &  6.44 &  $-0.81$ & 220 & 6.29 & $ 0.22$ & 41.11 &  ...           & ...     &  ...    &  ...    & $1.52$               &55,  9,  9,  9,          9 \\ 
 Kaz320                  & 0.034 & Sa         & $-17.87 $ & 1470 &  6.43 &  $-0.49$ & 260 & 6.59 & $ 0.56$ & 41.74 &  ...           & ...     &   ...   & ...     & $0.62\pm0.12$        &30,  1,  1, 47,          8 \\ 
 MCG+06-26-012           & 0.032 & SB0        & $-19.54 $ & 1685 &  6.83 &  $-0.39$ & 220 & 6.29 & $ ... $ &  ...  &  ...           & ...     &   ...   & ...     & $2.77\pm0.08$        &30, 30,  1,             30 \\ 
 MCG+08-23-067           & 0.030 & ...        & $  ...  $ &  730 &  5.64 &  $-0.02$ & 600 & 8.05 & $-1.44$ & 41.20 &  ...           & ...     &  ...    &  ...    & $1.38$               &     9,  9,  9,          9 \\ 
 MS2254-36               & 0.039 & ...        & $  ...  $ & 1530 &  6.58 &  $-0.43$ & 510 & 7.76 & $-0.39$ & 41.97 &  ...           & 0.19    &  0.58   & 0.76    & $1.78$               &    10,  9, 10,     11,  9 \\ 
 Mark1044                & 0.016 & SB0        & $-19.13 $ & 1280 &  6.50 &  $-0.22$ & 335 & 7.03 & $-0.48$ & 41.14 & 37.81$^{a}$    & 0.22    &  0.43   & 0.88    & $3.08\pm0.09$        &30, 30,  1,  5, 45, 13, 30 \\ 
 Mark110                 & 0.035 & disturbed  & $  ...  $ & 1760 &  6.87 &  $-0.42$ & ... & ...  & $  ...$ & 41.95 & 38.44          & ...     &   ...   & ...     & $2.35\pm0.05$        &    29, 29,  3,  6,     30 \\ 
 Mark1239$^{\ast}$       & 0.019 &  E-S0      & $-19.54 $ &  910 &  6.26 &  $ 0.12$ & 400 & 7.34 & $ 0.43$ & 42.36 & 38.42$^{b}$    & 1.14    &  1.33   & $<2.41$ & $2.94\pm0.14$        &30, 30,  1,  5, 25, 13, 30 \\ 
 Mark142                 & 0.045 & S?         & $-19.24 $ & 1620 &  6.87 &  $-0.29$ & 260 & 6.59 & $ 0.19$ & 41.37 & 37.60          & 0.10    &  0.15   & 0.80    & $3.15\pm0.11$        &30, 30,  1,  4, 38, 13, 30 \\ 
 Mark335$^{\ast}$        & 0.025 &  S0/a      & $-20.17 $ & 1640 &  7.07 &  $-0.16$ & 245 & 6.48 & $ 0.89$ & 41.97 & 37.91$^{b}$    & 0.38    &  0.34   & $<0.57$ & $3.10\pm0.05$        &30, 30,  1,  3, 25, 13, 30 \\ 
 Mark359                 & 0.017 & SB0a       & $-19.36 $ &  900 &  6.24 &  $ 0.13$ & 180 & 5.94 & $ 0.96$ & 41.50 & $<37.57$$^{b}$ & 0.44    &  1.13   & 1.74    & $2.40\pm0.10$        &     1,  1,  4, 25, 13, 30 \\ 
 Mark382                 & 0.034 & Sc         & $-17.65 $ & 1500 &  6.74 &  $-0.28$ & 155 & 5.68 & $ 1.52$ & 41.80 & ...            & $<0.24$ &  0.22   & $<0.74$ & $3.09\pm0.23$        &30, 30,  1,  4,     13, 30 \\ 
 Mark42                  & 0.024 & SBb        & $-18.28 $ &  670 &  5.89 &  $ 0.30$ & 220 & 6.29 & $-0.05$ & 40.84 & $<37.14$       & $<0.14$ &  0.32   & $<0.91$ & $2.76\pm0.23$        &30, 30,  1,  4, 16, 13, 30 \\ 
 Mark486                 & 0.038 & SBb?       & $-19.37 $ & 1480 &  6.94 &  $-0.10$ & 400 & 7.34 & $-0.32$ & 41.61 & $<37.82$       & 0.12    & $<0.13$ & $<0.57$ & $2.80\pm0.76$        &30, 30,  1,  3, 48, 13, 50 \\ 
 Mark493                 & 0.031 & SB(r)b     & $-18.84 $ &  740 &  6.17 &  $ 0.37$ & 315 & 6.92 & $-0.16$ & 41.36 & 37.86          & 0.19    &  0.69   & 1.29    & $2.84\pm0.14$        &30,  1,  1,  5, 16, 13, 30 \\ 
 Mark684                 & 0.046 & S?         & $-20.17 $ & 1400 &  7.00 &  $ 0.04$ & 170 & 5.84 & $ 1.39$ & 41.83 & $<37.71$       & 0.13    &  0.43   & 0.75    & $2.40\pm0.20$        &13, 30,  9,  5, 16  13, 30 \\ 
 Mark705                 & 0.028 & S0?        & $-19.75 $ & 1990 &  7.02 &  $-0.49$ & 360 & 7.15 & $-1.02$ & 40.73 & $<37.91$$^{b}$ & 0.22    &  0.59   & 0.92    & $2.33\pm0.09$        &30, 30,  1,  3, 23, 13, 30 \\ 
 Mark734                 & 0.049 & compact    & $  ...  $ & 1820 &  7.28 &  $-0.15$ & 315 & 6.92 & $ 0.15$ & 41.67 & $<37.77$       & 0.28    &  0.55   & 0.75    & $3.63\pm0.19$        &    30,  1,  5, 38, 13, 30 \\ 
 Mark739E                & 0.030 & ...        & $  ...  $ &  900 &  6.54 &  $ 0.36$ & 380 & 7.25 & $-0.38$ & 41.46 & ...            & 0.36    &  1.41   & 2.33    & $2.43\pm0.14$        &    30,  1,  3,      2, 30 \\ 
 Mkn766$^{\ast}$         & 0.013 & (R')SB(s)a & $-18.97 $ & 1100 &  6.28 &  $-0.16$ & 220 & 6.29 & $ 0.40$ & 41.29 & 37.05$^{b}$    & 1.30    &  4.03   & 4.66    & $2.79\pm0.11$        &30,  1,  1,  3, 16, 13, 30 \\ 
 Mark896                 & 0.027 & SBb        & $-19.04 $ & 1382 &  6.35 &  $ 0.21$ & 315 & 6.92 & $-0.03$ & 41.49 & $<37.98$$^{b}$ & 0.13    &  0.51   & 1.04    & $3.38\pm0.05$        &30,  3,  1,  5, 25, 13, 30 \\ 
 NGC4051$^{\ast}$        & 0.002 & SABbc      & $-16.60 $ &  990 &  5.58 &  $-0.55$ & 200 & 6.13 & $-0.41$ & 40.31 & 36.58          & 1.59    &  7.13   & 23.90   & $2.84\pm0.04$        & 4, 30,  1,  5, 17, 13, 30 \\ 
 NGC4748$^{\ast}$        & 0.014 & Sa         & $-18.73 $ & 1100 &  6.37 &  $-0.09$ & 295 & 6.81 & $ 0.26$ & 41.66 & $<37.31$$^{b}$ & 0.37    &  1.16   & 2.22    & $2.46\pm0.15$        &30, 30,  1, 12, 25, 13, 30 \\ 
 R14.01                  & 0.042 & ...        & $  ...  $ & 1790 &  7.19 &  $-0.19$ & 430 & 7.46 & $ 0.23$ & 42.29 & ...            & ...     &  ...    & ...     & ...                  &    30,  1,  3             \\ 
 RXJ1017.3+2914          & 0.049 & ...        & $  ...  $ & 1990 &  7.09 &  $-0.44$ & 255 & 6.55 & $ 0.70$ & 41.85 & ...            & 0.24    & 0.47    & 0.77    & $2.00\pm0.20$        &    10, 10, 10,     11, 11 \\ 
 RXJ1618.1+3619          & 0.034 & ...        & $  ...  $ &  830 &  5.95 &  $ 0.02$ & 100 & 4.92 & $ 1.17$ & 40.68 & ...            & 0.06    & 0.08    & 0.40    & $1.50\pm0.10$        &     9,  9,  9,     11,  9 \\ 
 RXJ2304.6-3501          & 0.042 & ...        & $  ...  $ & 1775 &  6.71 &  $-0.56$ & 210 & 6.21 & $ 1.26$ & 42.07 & ...            & 0.17    & 0.50    & $<0.42$ & $1.65$               &     9,  9,  9,     11,  9 \\ 
 WPVS007                 & 0.029 & ...        & $  ...  $ & 1620 &  6.85 &  $-0.31$ & 320 & 6.95 & $-0.15$ & 41.39 & ...            & ...     &  ...    &   ...   & $8.00\pm2.00$        &    10, 10, 10,         11 \\ 
 Z1136+3412              & 0.033 & SB0        & $-19.61 $ & 1450 &  6.32 &  $-0.55$ & 210 & 6.21 & $ 0.38$ & 41.19 & ...            & ...     &  ...    &   ...   & $1.90\pm0.20$        &30,  9,  9,  9,         11 \\ \hline 
\end{tabular}
\parbox{10.1in}
{\baselineskip 9pt
\noindent
{\sc Note}: 
(1): source name; (2): redshift; (3): Hubble type; (4):  bulge absolute magnitude in B band; (5): FWHM of $ \rm H \beta $ line (in $\rm km~s^{-1}$);					   
(6): mass of black hole  calculated with FWHM of $ \rm H \beta $; (7) Eddington ratio ($L_{\rm bol}=9 \times L_{5100\AA})$; (8): FWHM of \oiii line (in $\rm km~s^{-1}$); 
(9): mass of black hole  calculated with FWHM of \oiii line; 													          
(10) Eddington ratio ($L_{\rm bol}=3500\times L_{\rm [O~ III]})$ (11): luminosity of \oiii$\lambda 5007$ emission line; 					          
(12): luminosity of radio emission (5GHz); (13)-(15): infrared flux (in Jy) for $25\mu$m, $60\mu$m, $100\mu$m;  
(16): photon index of soft X-ray 0.1-2.4 keV; 0.2-2.0 keV (Grupe et al. 1998, 2004); 0.1-2.0 keV (Xu et al. 2003);
(17): reference (for column (4),(5),(8), (11), (12), (13)-(15), (16) respectively).
Objects marked with $^{\ast}$ is from 12$\mu$m sample, radio luminosity at 5 GHz marked with $a-e$ indicate that it is 
deduced from:
$^{a}$ --8.3GHz,
$^{b}$ --2.3GHz,
$^{c}$ --8.4GHz,
$^{d}$ --4.9GHz,
$^{e}$ --1.5GHz  						        	        

%\vglue -0.7cm
{\sc Reference}:
(1) Wang \& Lu 2001; (2) Dultzin-Hacyan et al. 1990; (3) Marziani 2003; (4) Whittle 1992; (5) Dahari 1988;  
(6) Ulvestad \& Wilson 1984(a); (7) Stepanian et al. 2003; (8) Xu et al. 2003; (9) Grupe et al. 2004;
(10) Grupe et al. 1999; (11) Grupe et al. 1998(b);  (12) de Grijp 1992; (13) NED; (14) Kraemer 2004. (15) Tran 2003 ; 
(16) Ulvestad et al. 1995; (17) Ho \& Ulvestad 2001; (18) Shu et al. 2006; (19) Bassani et al. 1999; (20) Lumsden et al. 2004; 
(21) Lumsden et al. 2001; (22) Gu \& Huang 2002; (23) Roy et al. 1998; (24) Nelson \& Whittle 1995; (25) Roy et al. 1994; 
(26) Sadler 1995; (27) Thean et al. 2000; (28) Heisler et al. 1998;(29) Vestergaard, 2002; (30) Veron-Cetty, et al. 2001;
(31) Gelderman \& Whittle 1994; (32) Shier \& Fischer, 1998; (33) Kim 1995; (34) Keel 1983; (35) Wilson \& Baldwin 1989; 
(36) Yong, et al 1996; (37) Ulvestad \& Wilson 1981; (38) Kerllermann et al. 1989; (39) Ulvestad \& Wilson 1989;
(40) Nager et al. 1999; (41) Ulvestad \& Wilson 1984(b); (42) Acker et al. 1991; (43) Storchi-Bergmann et al. 1995; 
(44) Durret \& Bergeron 1988; (45) Kinney et al. 2000; (46) Schmitt et al. 1997 (b); (47) Zamorano et al. 1992; 
(48) Lonsdale et al. 1995; (49) Boller et al. 1992; (50) Pfefferkorn et al. 2001;
(51) Detuit 2004; (52) Winkler 1992; (53) Rush et al. 1996; (54) Duric et al. 1983; (55)Grazian et al. 2000}
\end{center}          
\end{landscape}

\clearpage

\begin{landscape}

\begin{center}{\sc \small Table 3. The Non-hidden Broad Line Seyfert 2 Sample}
\footnotesize
\vglue 0.2cm
\begin{tabular}{lccccccccccccl}\hline\hline
Name&$z$& morphology&$M_{\rm bulge}$ &FWHM  &$\log\mbh$ &$\log \cal{E}$&$\log L_{\rm [O~ III]}$ &$\log L_{\rm 5GHz}$ &$f_{25}$&$f_{60}$&$f_{100}$ &$\log_{10}N_{\rm H}$ & Ref. \\ 
    &   &               &            & \oiii&            &                  &                       &                    &        &        &          &  &        \\    
(1) &(2)&    (3)         &  (4) &  (5)       &      (6)         &          (7)          &         (8)        &   (9)  &  (10)  &  (11)    &  (12)&  (13)&  (14)   \\ \hline                                                      
ESO428-G014          & 0.006 & SA0          & $-18.03 $ & 400 & 7.34  & $ 0.29 $ & 42.23 & 37.99          & 1.77  &  4.40 &   6.05  &$>25.00$&13, 35, 42, 46, 13, 22 \\
IC5298               & 0.027 & ...          & $  ...  $ & ... & ...   & $  ... $ & 42.17 &  ...           & 1.80  &  9.76 &  11.10  &   ...  &        22,     13     \\
F03362-1642$^{\ast}$ & 0.037 & SBb          & $-18.93 $ & ... & ...   & $  ... $ & 41.64 & $<38.16$$^{b}$ & 0.50  & 1.06  & $<2.01$ &   ...  &13,     12, 25, 13     \\
IRAS04210+0400       & 0.045 & S?           & $-18.48 $ & 400 & 7.34  & $ 0.48 $ & 42.42 &  ...           & 0.25  &  0.60 & $<2.54$ &   ...  &13, 31, 22,     13     \\ 
IRAS04229-2528       & 0.044 & S0/a         & $-19.99 $ & ... & ...   & $  ... $ & 41.99 & $<38.32$$^{b}$ & 0.26  &  0.98 &   1.25  &   ...  &13,     22, 25, 13     \\ 
IRAS08277-0242       & 0.041 & SB(rs)b      & $-19.25 $ & ... & ...   & $  ... $ & 41.76 & $<38.56$$^{b}$ & 0.43  &  1.47 &   1.82  &   ...  &13,     12, 25, 13     \\ 
IRAS10340+0609       & 0.012 & ...          & $  ...  $ & ... & ...   & $  ... $ &  ...  & 36.74          &$<0.25$&  0.39 & $<1.12$ &   ...  &            39, 13     \\ 
IRAS13452-4155       & 0.039 & ...          & $  ...  $ & 250 & 6.52  & $ 1.07 $ & 42.19 & 38.34$^{b}$    & 0.81  &  1.84 &   1.34  &   ...  &    36, 22, 28, 13     \\ 
IRAS23128-5919       & 0.045 & ...          & $  ...  $ & ... & ...   & $  ... $ &  ...  &  ...           & 1.59  & 10.80 &  11.00  &   ...  &                13     \\ 
M51$^{\ast}$         & 0.002 & SA(s)bc      & $-18.95 $ & 190 & 6.04  & $-0.55 $ & 40.09 & 35.53          & 17.5  &108.70 &  292.10 & 23.88  &13, 24, 15, 17, 15, 15 \\ 
Mrk266SW             & 0.028 & pec,dble     & $  ...  $ & 315 & 6.92  & $ 0.34 $ & 41.86 & 39.34$^{c}$    & 1.13  & 7.27  &  10.07  & 25.00  & 4,  4, 15, 27, 15, 15 \\
Mrk573               & 0.017 & (R)SAB(rs)0  & $-19.72 $ & 190 & 6.04  & $ 1.75 $ & 42.39 & $<37.72$$^{b}$ & 0.81  & 3.60  &  1.30   &$>24.00$& 4,  4, 15, 25, 15, 18 \\
Mrk938$^{\ast}$      & 0.020 & Sc           & $-16.74 $ & 330 & 7.00  & $ 1.09 $ & 42.69 & 39.03$^{c}$    & 2.51  & 16.84 &  17.61  & 23.00  &13,  5,  5, 27, 15, 15 \\
Mrk1066              & 0.012 & (R)SB(s)0    & $-19.51 $ & 440 & 7.50  & $ 0.17 $ & 42.27 & 38.11$^{a}$    & 2.26  & 11.00 &  12.20  &$>24.00$& 4,  4, 22, 40, 13, 22 \\
Mrk1361              & 0.023 & SB           & $-18.62 $ & ... & ...   & $  ... $ & 42.33 & 38.34$^{c}$    & 0.84  & 3.28  &  3.73   &   ...  &13,     22, 27, 13     \\
NGC1144$^{\ast}$     & 0.029 & S pec        & $-20.54 $ & 170 & 5.84  & $ 1.37 $ & 41.81 & 38.07$^{b}$    & 0.62  & 5.35  &  11.60  & 22.00  &13, 33, 15, 25, 15, 15 \\
NGC1241$^{\ast}$     & 0.014 & SB(rs)b      & $-19.37 $ & 400 &$<7.34$& $<-0.20$ & 41.74 &  ...           & 0.60  & 4.37  &  10.74  &   ...  &13,  5, 15,     15     \\
NGC1320$^{\ast}$     & 0.009 & Sa           & $-18.45 $ & 229 & 6.36  & $ 0.12 $ & 41.08 & 37.22$^{c}$    & 1.32  & 2.21  &  2.82   &   ...  & 4, 24, 15, 27, 15     \\
NGC1358              & 0.013 & SAB(r)0/a    & $-19.73 $ & 220 & 6.29  & $ 0.47 $ & 41.36 & 37.32          &$<0.12$&  0.38 &   0.93  &   ...  & 4,  4,  4, 39, 13     \\ 
NGC1386$^{\ast}$     & 0.003 & SB(s)0       & $-17.78 $ & 315 & 6.92  & $-0.43 $ & 41.09 & 37.05          & 1.46  & 6.01  &  9.67   & 25.00  & 4, 24, 15, 41, 15, 15 \\
NGC1667$^{\ast}$     & 0.015 & SAB(r)c      & $-18.61 $ & 275 & 6.68  & $ 0.75 $ & 42.03 & 37.82$^{c}$    & 0.67  & 6.29  &  15.83  & 24.00  & 4,  4, 43, 27, 15, 15 \\
NGC1685              & 0.015 & SB(r)0/a     & $-18.54 $ & ... & ...   & $  ... $ & 42.67 & 38.06          & 0.22  &  0.98 &   1.53  &   ...  &13,     22, 39, 13     \\ 
NGC3079$^{\ast}$     & 0.004 & SB(s)c       & $-17.51 $ & ... & ...   & $  ... $ & 40.48 & 38.07$^{d}$    & 3.65  & 50.95 &  105.20 & 22.20  &13,     15, 54, 15, 15 \\
NGC3281              & 0.012 & SAB(rs+)a    & $-19.56 $ & 235 & 6.41  & $ 0.29 $ & 41.30 & 38.66$^{a}$    & 2.63  &  6.86 &   7.51  & 23.90  & 4,  5, 44, 45, 13, 19 \\ 
NGC3362              & 0.028 & SABc         & $-19.02 $ & 369 & 7.20  & $-0.43 $ & 41.37 & $<38.13$$^{b}$ & 0.35  & 2.13  &  3.16   &   ...  &13, 24, 15, 25, 15     \\
NGC3393              & 0.013 & (R')SB(s)ab  & $-19.12 $ & ... & ...   & $  ... $ & 42.10 & 37.98$^{b}$    & 0.75  & 2.25  &  3.87   &$<23.85$&13,     19, 25, 13, 19 \\  
NGC3982$^{\ast}$     & 0.004 & SAB(r)b      & $-17.70 $ & 203 & 6.15  & $-0.42 $ & 40.33 & 36.31$^{c}$    & 0.97  & 7.21  &  16.78  &$>24.20$&13, 24, 15, 27, 15, 18 \\
NGC4117              & 0.003 & S0           & $-16.08 $ & 189 & 6.03  & $  ... $ &  ...  & 35.71          &  ...  &  ...  &   ...   &   ...  &13, 24,     46         \\
NGC4941$^{\ast}$     & 0.004 & (R)SAB(r)ab  & $-17.89 $ & 226 & 6.34  & $ 0.24 $ & 41.18 & 37.16$^{c}$    & 0.46  & 1.87  &  4.79   & 23.65  &13, 24, 15, 27, 15, 15 \\
NGC5128              & 0.002 & S0 pec       & $-21.06 $ & ... & ...   & $  ... $ & 38.82 & ...            & 28.2  & 213.0 &  412.0  &$>23.00$&13,     19,     13, 19 \\  
NGC5135$^{\ast}$     & 0.014 & SB(l)ab      & $-19.60 $ & 165 & 5.79  & $ 1.82 $ & 42.21 & $<37.91$$^{c}$ & 2.39  & 16.60 &  31.18  &$>24.00$& 4,  4,  4, 27, 15, 22 \\
NGC5283              & 0.010 & S0           & $-18.28 $ & 358 & 7.14  & $-0.86 $ & 40.88 & 37.74          & 0.13  & 0.21  &  0.27   & 23.18  & 4,  5, 15, 46, 15, 18 \\
NGC5347$^{\ast}$     & 0.008 & (R')SB(rs)ab & $-17.75 $ & 392 & 7.30  & $-0.68 $ & 41.22 & 37.13          & 0.96  & 1.42  &  2.64   & 24.00  &13, 24, 15, 39, 15, 15 \\
NGC5643              & 0.004 & SAB(rs)c     & $-17.78 $ & 240 & 6.45  & $ 0.32 $ & 41.37 & 37.49          & 3.65  & 19.50 &  38.20  & 23.85  & 4,  4,  5, 46, 13, 18 \\  
NGC5695              & 0.014 & SBb          & $-18.63 $ & 359 & 7.15  & $-1.20 $ & 40.55 & $<36.98$       & 0.13  & 0.566 &  1.79   &   ...  &13, 24, 15, 39, 15     \\
NGC5728              & 0.009 & (R1)SAB(r)a  & $-19.45 $ & 320 & 6.95  & $ 0.42 $ & 41.09 & 37.55          & 0.88  &  8.16 &  14.70  &   ...  & 4,  4,  5, 39, 13     \\ 
NGC6300              & 0.004 & SBb          & $-18.35 $ & 220 & 6.29  & $ 0.19 $ & 41.08 & $<37.33$       & 2.27  & 14.70 &  36.00  & 23.34  & 4,  4, 20, 26, 13, 18 \\ 
NGC6890$^{\ast}$     & 0.008 & (R')SA(r)ab  & $-18.09 $ & 245 & 6.48  & $-0.04 $ & 41.04 & 37.27$^{c}$    & 0.65  & 3.85  &  8.16   &   ...  & 4,  4, 15, 27, 13     \\
NGC7172$^{\ast}$     & 0.009 & Sa           & $-18.93 $ & ... & ...   & $  ... $ & 40.84 & 37.83$^{c}$    & 0.95  & 5.74  &  12.43  & 22.94  &13,     15, 27, 15, 15 \\
NGC7582$^{\ast}$     & 0.005 & (R'1)SB(s)ab & $-18.87 $ & 186 & 5.99  & $ 1.04 $ & 41.63 & 38.27          & 7.48  & 52.47 &  83.27  & 23.09  & 4,  4, 15, 41, 15, 15 \\
NGC7672              & 0.013 & Sb           & $-17.34 $ & 297 & 6.82  & $  ... $ &  ...  & 36.85$^{e}$    &$<0.15$&  0.46 & $<2.46$ &  ...   &13, 24,     41, 13     \\ 
UGC6100              & 0.030 & Sa           & $-20.10 $ & 679 & 8.26  & $-0.56 $ & 42.30 & 38.13$^{a}$    & 0.28  & 0.81  &  1.96   &  ...   &13, 24, 15, 45, 15     \\ \hline  	        
\end{tabular}
\parbox{8.5in}
{\baselineskip 9pt
\noindent
{\sc Note}:
{(1): source name; (2): redshift; (3):  Hubble type; (4): bulge absolute magnitude in B band;  
(5): FWHM of \oiii line (in $\rm km~s^{-1}$); (6): mass of black hole; (7) Eddington ratio; 
(8): luminosity of \oiii $\lambda 5007$ emission line; (9): luminosity of radio emission (5GHz); 
(10)-(12): infrared flux (in Jy) for $25\mu$m, $60\mu$m, $100\mu$m; (13)Column density ;(14): reference 
(for column (4),(5),(7),(8),(9), (10)-(12),(13) respectively).
Objects marked with $^{\ast}$ are from 12$\mu$m sample, radio luminosity at 5 GHz marked with $a-e$ indicate that it is 
deduced from:
$^{a}$ --8.3GHz,
$^{b}$ --2.3GHz,
$^{c}$ --8.4GHz,
$^{d}$ --4.9GHz,
$^{e}$ --1.5GHz}   
\vglue -0.4cm 																						       
}          
\end{center}
          
\end{landscape}

\end{document}